\documentclass[10pt,aps,pra,showkeys,twocolumn,superscriptaddress]{revtex4-2}
\usepackage{amsmath, amssymb}
\usepackage{mathrsfs}
\usepackage{array}
\usepackage{booktabs}
\usepackage{tabu}
\usepackage{dcolumn}
\usepackage{amsmath}
\usepackage{amsfonts}
\usepackage{float}
\usepackage{amssymb}
\usepackage{graphicx,color}
\usepackage[colorlinks={true}]{hyperref}
\usepackage{multirow, makecell}

\hypersetup{colorlinks=true,linkcolor=red,citecolor=blue,urlcolor=blue}
\usepackage{xcolor}
\usepackage{comment}
\usepackage{subcaption}
\usepackage{caption}
\usepackage{graphicx}
\usepackage{graphicx}
\usepackage{dcolumn}
\usepackage{bm}
\usepackage{pstricks}
\usepackage{braket}
\usepackage{orcidlink}
\usepackage{tikz}
\def\be{\begin{equation}}
  \def\ee{\end{equation}}
\def\bea{\begin{eqnarray}}
\def\eea{\end{eqnarray}}
\def\f{\frac}
\def\n{\nonumber}
\def\l{\label}
\def\p{\phi}
\def\o{\over}
\def\R{\rho}
\def\pa{\partial}
\def\om{\omega}
\def\na{\nabla}
\def\P{\Phi}

\usepackage{booktabs} 

\begin{document}

\title{Entanglement-enhanced optimal quantum metrology}
\author{Muhammad Talha Rahim \orcidlink{0000-0001-9243-417X}} \email{mura68827@hbku.edu.qa}
\affiliation{Qatar Centre for Quantum Computing, College of Science and Engineering, Hamad Bin Khalifa University, Doha, Qatar}
\author{Saif Al-Kuwari \orcidlink{0000-0002-4402-7710}} \email{smalkuwari@hbku.edu.qa}
\affiliation{Qatar Centre for Quantum Computing, College of Science and Engineering, Hamad Bin Khalifa University, Doha, Qatar}
\author{Asad Ali\orcidlink{0000-0001-9243-417X}} \email{asal68826@hbku.edu.qa}
\affiliation{Qatar Centre for Quantum Computing, College of Science and Engineering, Hamad Bin Khalifa University, Doha, Qatar}
\date{\today}
\def\be{\begin{equation}}
\def\ee{\end{equation}}
\def\bea{\begin{eqnarray}}
\def\eea{\end{eqnarray}}
\def\f{\frac}
\def\n{\nonumber}
\def\l{\label}
\def\p{\phi}
\def\o{\over}
\def\R{\rho}
\def\pa{\partial}
\def\om{\omega}
\def\na{\nabla}
\def\P{$\Phi$}

\begin{abstract}
Quantum optimal control (QOC) schemes can be employed to enhance the sensitivity of quantum metrology (QM) protocols undergoing Markovian noise, which can limit their precision to a standard quantum limit (SQL)-like scaling. In this paper, we propose a QOC scheme for QM that leverages entanglement and optimized coupling interactions with an ancillary system to provide enhanced metrological performance under general Markovian dynamics. We perform a comparative analysis of our entanglement-enhanced scheme against the unentangled scheme conventionally employed in QOC-enabled QM for varying evolution times and decoherence levels, revealing that the entanglement-enhanced scheme enables significantly better noise performance, even when a noisy ancilla is employed. We further extend our investigation to time-inhomogeneous noise models, specifically focusing on a noisy frequency estimation scenario within a spin-boson bath, and evaluate the protocol's performance under completely dissipative and dephasing dynamics. Our findings indicate that, in certain situations, schemes employing coherent control of a single particle are severely limited. In such cases, employing the entanglement-enhanced scheme can provide improved performance.
\end{abstract}


\maketitle


\section{Introduction}\label{sec:intro}
The stochastic nature of quantum mechanics limits the achievable precision in estimating the physical parameters. When utilizing classical probes, the precision of the estimand measured after averaging the outcome is limited by the standard quantum limit (SQL) \cite{GLM:04:Science,GLM:06:PRL,GLM:11:Nature,TA:14:JPA}, which is a consequence of the central limit theorem. Quantum metrology (QM) protocols employ probes that use quantum properties to deliver an advantage in precision measurement over classical strategies. Quantum effects such as superposition 
\cite{GLM:04:Science}, squeezing \cite{LSOS:20:Quantum}, and entanglement \cite{GLM:06:PRL,GLM:11:Nature,TA:14:JPA} have been used to surpass the SQL and achieve a more fundamental limit in precision measurement, known as the Heisenberg limit (HL) \cite{GLM:04:Science,GLM:06:PRL,GLM:11:Nature,TA:14:JPA,LSOS:20:Quantum,KD:13:NJP,TA:14:JPA, DM:14:PRL,SSK:17:Quantum,ZZP:18:Nature,KGAD:23:PRL,CBM:13:PRL,ARTG:18:Quantum,IYHP:22:PRX,CHP:12:PRL,AY:21:PRL,YLLT:24:Nature,LY:17:PRA,PJ:17:Nature,XLL:19:NJPQI, YPC:22:PRL,ZYKX:23:PRA,NBCA:16:PRA,WWZB:18:PRA,KGAD:23:PRL}.

As a consequence of the HL, it is established that the scaling of the precision does not go beyond $\frac{1}{T}$, where $T$ is the total probing time. In ideal quantum systems, multipartite entanglement or enhanced interrogation times of quantum states utilizing quantum coherence can reduce the estimation error and help us obtain the HL in precision \cite{GLM:04:Science,GLM:06:PRL,GLM:11:Nature,TA:14:JPA}. However, such strategies rely on the unitarity of the time evolution and, in the presence of environmental decoherence effects, suffer a decline in their achievable precision \cite{KD:13:NJP,TA:14:JPA, DM:14:PRL}. Under the influence of the general class of Markovian noise channels, the achievable precision is reduced to a constant factor rather than a quadratic improvement over the SQL if the Hamiltonian not in Lindblad span (HNLS) condition is not met \cite{SSK:17:Quantum,ZZP:18:Nature}. However, the constant factor improvement still proves vital, especially in the case of limited resources, which is more experimentally relevant \cite{SSK:17:Quantum, KGAD:23:PRL}. 
%
%
In this case, noise mitigation techniques have been proposed to enhance the achievable sensitivity of sensing protocols. Popular noise mitigation strategies include optimizing the probing time for multiqubit systems \cite{CBM:13:PRL}, monitoring the environment \cite{ARTG:18:Quantum,IYHP:22:PRX}, exploiting non-Markovian effects \cite{CHP:12:PRL,AY:21:PRL,YLLT:24:Nature}, and quantum error correction(QEC) \cite{DSF:14:PRL,ZZP:18:Nature,SMM:21:NJP,CBBEH:23:RMP}. Such strategies, however, have to be tailored for specific noise processes or require entanglement between arbitrarily large quantum systems. Quantum optimal control (QOC) is an alternative strategy to enhance QM in the presence of noise. By leveraging precisely tailored electromagnetic pulses, QOC has enabled more efficient and practically realizable QM protocols for single and multi-parameter estimation experiments \cite{DCS:17:PRX,LY:17:PRA,PJ:17:Nature,SSK:17:Quantum,XLL:19:NJPQI, YPC:22:PRL,ZYKX:23:PRA}. 

Single-parameter QM protocols utilizing QOC typically rely on single-particle sequential encoding schemes  \cite{LY:17:PRA,XLL:19:NJPQI, ZYKX:23:PRA} with the potential benefits of entanglement in these protocols still underexplored. Entangling the probe with ancillary systems is a well-established strategy to enhance precision under finite resource constraints \cite{DM:14:PRL, HMM:16:PRA,NBCA:16:PRA,WWZB:18:PRA,KGAD:23:PRL}, or to enable QEC-based QM protocols \cite{SSK:17:Quantum,ZZP:18:Nature,CBBEH:23:RMP}. In this paper, we will study a QOC-enabled QM protocol that can leverage entanglement with an ancillary system as a resource to provide enhanced performance.

\subsection{Contribution}
We propose an entanglement-enhanced strategy employing QOC, where the probe is entangled with a single auxiliary qubit, and quantum control operations are applied to the combined system at finite intervals. We conduct a comparative analysis between the QOC-enabled unentangled strategy and our entanglement-enhanced setup, exploring both noiseless and noisy ancilla scenarios. Our analysis focuses on phase-covariant (PC) dynamics  \cite{SKHD:16:PRL, HSK:18:NJP}, specifically examining its time-homogeneous and time-inhomogeneous subtypes. 
For the case of time-homogeneous dynamics, we demonstrate that our strategy provides greater robustness to noise by evaluating its performance against the unentangled scheme for varying probing times and decay parameters. Notably, even with a noisy ancilla, the performance surpasses that of single-particle schemes. Additionally, we show how the superior performance of the entanglement-enhanced scheme yields controls that are more resilient to parameter variations. For the case of time-inhomogeneous dynamics, we explore the performance of QOC for frequency estimation utilizing a spin-$\frac{1}{2}$ system coupled to a bosonic bath. We cover both fully dissipative and dephasing dynamics and demonstrate the efficacy of our entanglement-enhanced scheme.

\subsection{Organization}
The rest of this paper is organized as follows: Sec. \ref{sec:prelim} introduces key definitions relevant to QOC and noisy QM. In Sec. \ref{sec:fw}, we outline the framework for applying our control mechanism and the associated constraints. Sec. \ref{sec:app} presents the sensitivity performance of the proposed protocols under various types of time-homogeneous and time-inhomogeneous noise. Finally, Sec. \ref{sec:dis} concludes the paper and provides an outlook for future work.

\begin{table*}[ht]
    \centering
    \begin{tabular}{|>{\centering\arraybackslash}p{3cm}|>{\centering\arraybackslash}p{3cm}|>{\centering\arraybackslash}p{4cm}|>{\centering\arraybackslash}p{6cm}|}
        \hline
        \textbf{} & \textbf{Abbreviation} & \textbf{Expansion} & \textbf{Description} \\ \hline
        \multirow{3}{*}{\raisebox{-20pt}[0pt][0pt]{\shortstack{\textbf{Uncontrolled} \\ \textbf{Schemes}}}} 
                                & \raisebox{-5pt}[0pt][0pt]{\centering UE}       & \raisebox{-5pt}[0pt][0pt]{\centering Unentangled}       & Unentangled sensing qubit undergoing sequential encoding \\ \cline{2-4}
                                & \raisebox{-5pt}[0pt][0pt]{\centering NLA}      & \raisebox{-5pt}[0pt][0pt]{\centering Noiseless Ancilla} & Sensing qubit entangled with noiseless ancilla undergoing sequential encoding \\ \cline{2-4}
                                & \raisebox{-5pt}[0pt][0pt]{\centering NA}       & \raisebox{-5pt}[0pt][0pt]{\centering Noisy Ancilla}     & Sensing qubit entangled with noisy ancilla undergoing sequential encoding \\ \hline
        \multirow{3}{*}{\raisebox{-30pt}[0pt][0pt]{\shortstack{\textbf{Controlled} \\ \textbf{Schemes}}}} 
                                & \raisebox{-5pt}[0pt][0pt]{\centering CUE}      & \raisebox{-5pt}[0pt][0pt]{\centering Controlled Unentangled} & Unentangled sensing qubit undergoing sequential encoding and assisted by QOC \\ \cline{2-4}
                                & \raisebox{-10pt}[0pt][0pt]{\centering CNLA}     & \raisebox{-10pt}[0pt][0pt]{\centering Controlled Noiseless Ancilla} & Sensing qubit entangled with noiseless ancilla undergoing sequential encoding and assisted by QOC \\ \cline{2-4}
                                & \raisebox{-10pt}[0pt][0pt]{\centering CNA}      & \raisebox{-10pt}[0pt][0pt]{\centering Controlled Noisy Ancilla} & Sensing qubit entangled with noisy ancilla undergoing sequential encoding and assisted by QOC \\ \hline
    \end{tabular}
    \caption{The QM schemes analyzed in this paper. The schemes are divided into uncontrolled schemes which do not employ QOC and controlled schemes which are assisted by QOC.}
    \label{tab:tab1} 
\end{table*}

\section{Preliminaries} \label{sec:prelim}
In this section, we present some basic definitions relevant to QM for open quantum systems described by Markovian dynamics. We also provide concise discussions on concepts related to QOC.

\subsection{Quantum Metrology}\label{sec:subsec}

A QM experiment generally consists of four key steps: (i) the preparation of a suitable probe state, (ii) its interaction with the system of interest, (iii) measurement using the optimal positive operator-valued measure (POVM), and (iv) post-processing to estimate the parameter of interest, $\omega_0$. The POVM is a set $\{M_x\}$ where the elements are positive semidefinite operators and satisfy $\sum_x M_x = 1$. The measurement operators can correspond to either local or global measurements of the system.

The minimal mean square error (MSE) attainable by any consistent and unbiased estimator is governed by the Cramér-Rao bound (CRB):

\begin{align}\label{eq: eq_1}
\Delta \tilde{\omega}_0 \geq \frac{1}{\sqrt{\mathcal{F}_{\text{cl}}[p_{\omega_0}(t)]}},
\end{align}

\noindent where

\begin{align} \label{eq: eq_2}
\mathcal{F}_{\text{cl}}[p_{\omega_0}] = \sum_x \frac{\dot{p}_{\omega_0}(x)^2}{p_{\omega_0}(x)},
\end{align}

\noindent is the classical Fisher information (CFI), defined by the probability distribution $p_{\omega_0}(x)$, which depends on the system parameter $\omega_0$. Since $p_{\omega_0} = \text{Tr}[\rho_{\omega_0}(t)M_x]$, the Cramér-Rao bound sets the limit on the attainable precision of the protocol using the using the measurement operators, $\{M_x\}$.

The quantum Fisher information (QFI) is the quantity in \ref{eq: eq_2} optimized over all measurements, and is given by:

\begin{align}\label{eq: eq_3}
\mathcal{F}_Q[\rho_{\omega_0}(t)] =  \text{Tr}[\rho_{\omega_0}(t) L_{\omega_0}^2],
\end{align}

\noindent which is obtained by maximizing $\mathcal{F}_{cl}\left[\rho_{\omega_0}(t) \right]$ over the span of $\{M_x\}$. $L_{\omega_0}$ represents the symmetric logarithmic derivative (SLD) that satisfies

\begin{align}\label{eq: eq_4}
\dot{\rho}_{\omega_0}(t) = \frac{1}{2} \left( \rho_{\omega_0}(t)L_{\omega_0} + L_{\omega_0}\rho_{\omega_0}(t) \right).
\end{align}

The evaluation of the SLD generally depends on the eigendecomposition of $\rho_{\omega_0}(t)$, which becomes straightforward when the condition of unitarity of the overall system holds. Under this scenario, the system evolution is dictated by the parameter-encoding Hamiltonian, $H_{\omega_0}$ represented by the superoperator $\mathcal{U}_{\omega_0} = e^{-\iota \omega_0 H} \boldsymbol{\cdot} e^{\iota \omega_0 H}$. However, the more general case is the noisy evolution given by

\begin{align}\label{eq:eq_5}
\rho_{\omega_0}(t) =  \Lambda_{\omega_0}\left(t\right)\rho(0) 
\end{align}

\noindent where $\Lambda_{\omega_0}\left(t\right) = \mathcal{U}_{\omega_0} \left(t\right) \circ \Gamma \left(t \right)$ where the superoperator $\Gamma \left(t \right)$ represents the noise. In the case of PC channels considered in this paper, $\mathcal{U}_{\omega_0}\left(t\right) \circ \Gamma \left(t \right)=\Gamma \left(t \right) 
\circ \mathcal{U}_{\omega_0}\left(t\right)$.

The efficacy of the QM protocol can thus be quantified by the quantum Cramér-Rao Bound (QCRB), which can be described by:

\begin{align}
\Delta \tilde{\omega}_0 \geq \frac{1}{\sqrt{v \mathcal{F}_Q[\rho_{\omega_0}(t)]}},
\end{align}

\noindent where $v$ denotes the number of repetitions of the protocol.

\subsection{Quantum Optimal Control}
Comprehensive control of a quantum system is achieved through the application of external electromagnetic fields, resulting in a system Hamiltonian of the form:

\begin{equation} \label{eq: control}
H = H_{\omega_0} + H_c,
\end{equation}

\noindent where $H_{\omega_0}$ represents the free evolution Hamiltonian dependent $\omega_0$, and $H_c$ represents the control Hamiltonians, induced by the interaction with the external fields.

These control fields become particularly significant when considering noisy interactions. The short-time expansion of the dynamics in \ref{eq:eq_5} is described by the master equation:

\begin{equation} \label{eq: lind_1}
\dot{\rho}_{\omega_0} = \mathcal{L}[\rho_{\omega_0}],
\end{equation}

\noindent where $\mathcal{L}$ is the Liouvillian superoperator, and $\dot{\rho}(t) = \frac{d}{d\omega}\rho_{\omega_0}(t)$. This can be further simplified into a combination of free evolution under the system Hamiltonian, $H$, and a decoherence term acting on the state:

\begin{align}\label{eq: lind_2}
\dot{\rho}(t) &= -i[H, \rho_{\omega_0}] + \nonumber \\
&\quad \sum_i \gamma_i(t) \left( L_i \rho_{\omega_0} L_i^\dagger - \frac{1}{2} \{\rho_{\omega_0}, L_i^\dagger L_i\} \right),
\end{align}

\noindent where $L_i$ are the Lindblad operators. The decoherence rate, $\gamma_i(t)$, represents the intensity of the noise encountered by the system and can be either time-dependent or time-independent.

We assume Markovian dynamics throughout this paper. This assumption holds when the environment’s correlations decay significantly faster than the system’s evolution \cite{MM:22:Quantum}. Such dynamics are described as memoryless, meaning that noise applied at a time $t_1$ does not depend on noise applied at earlier times. Generally, the total evolution time satisfies $T \gg \Delta t$, and our control steps are incorporated within these time steps. The evolved final state after time $T$ over multiple control steps can be expressed as:

\begin{equation} \label{eq: lind_exp}
\rho(T) = \prod_{i=1}^m e^{\Delta t \mathcal{L}_i},
\end{equation}

\noindent where $m$ is the number of time steps. In this framework, we account for both time-dependent and time-independent decoherence, corresponding to whether the decoherence rate changes over time. Time inhomogeneity introduces additional complexity to the optimization process, requiring the evolution to adapt to varying decoherence rates over time. For this case, we assume that $\mathcal{L}_i$ remains constant over each timestep $\Delta t$ but changes for every $i$.

\begin{figure*}[t]      
  \centering
  \begin{tikzpicture}
      \node[anchor=center] (img2) {\includegraphics[width=0.8\textwidth]{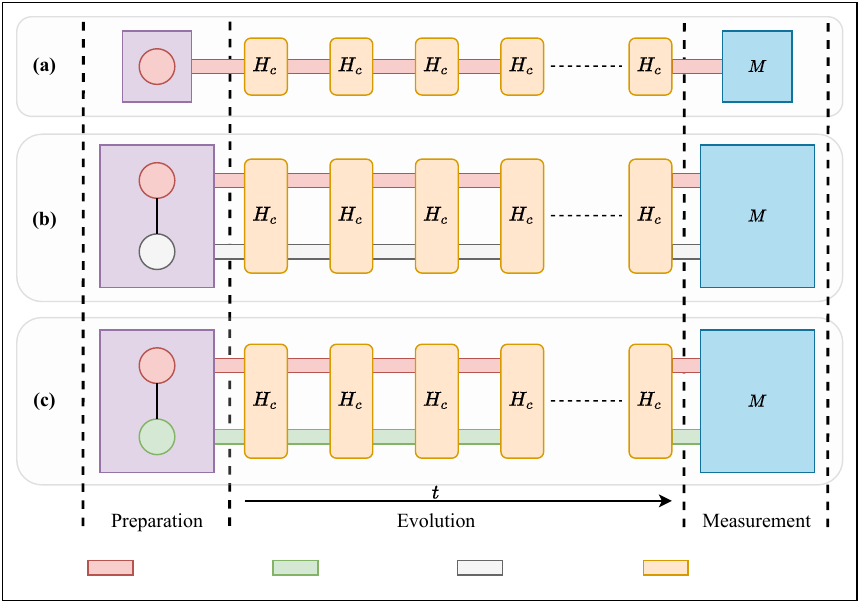}};
      \node at (-3.85, -4.30){$\omega$-dependent };
     \node at (-3.85, -4.70){noisy evolution };
      \node at (-0.75, -4.30){$\omega$-dependent };
     \node at (-0.75, -4.70){noisy evolution };
     \node at (2.15, -4.30){Noiseless};
    \node at (2.15, -4.70){evolution};
    \node at (5.45, -4.40){Controls};

  \end{tikzpicture}

  \caption{\raggedright QM protocols employing the \textbf{(a)} CUE scheme, \textbf{(b)} CNLA scheme, and \textbf{(c)} the CNA scheme. For the CNLA and CNA cases, control operations take place on both particles.}
  \label{fig:schemes}
\end{figure*}
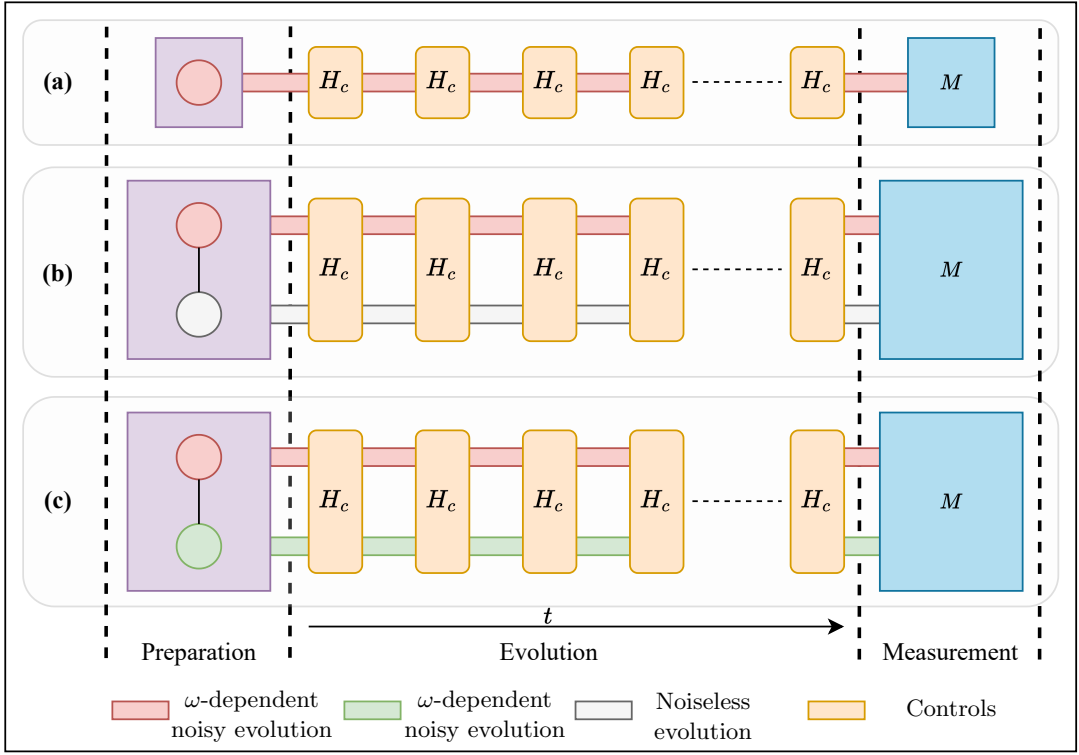

\section{Framework}\label{sec:fw}
In this paper, we apply QOC to three distinct QM protocols:
i) the \textbf{controlled unentangled (CUE)} scheme, involving a single particle,
ii) the \textbf{controlled noiseless ancilla (CNLA)} scheme, where the sensing qubit is entangled with a noiseless ancilla, and
iii) the \textbf{controlled noisy ancilla (CNA)} scheme, where the sensing qubit is entangled with a noisy ancilla. 
We will also examine the uncontrolled counterparts of these protocols, namely the \textbf{unentangled (UE)}, \textbf{noiseless ancilla (NLA)}, and \textbf{noisy ancilla (NA)} schemes for a more comprehensive analysis. The schemes are summarized in Table \ref{tab:tab1}. In this context, we use the Gradient Ascent Pulse Engineering (GRAPE) algorithm \cite{KRKH:05:JMR, LY:17:PRA} assisted with auto differentiation (AD), which provides faster computation against ordinary GRAPE \cite{ZYYW:22:PRR} and will be more efficient than non-gradient based algorithms for larger systems.

For the single-particle case, we define the control Hamiltonian as:

\begin{align*}
H_c = \sum_{i=1}^3 u_i(t) \sigma_i,
\end{align*}

\noindent where \(\sigma_i\) are the local Pauli matrices acting on the qubit, and \(u_i(t)\) are the corresponding control coefficients. This allows us to apply control over the single qubit using a combination of Pauli operators.

In the case of the entanglement-enhanced system, we assume full control over both particles. The control Hamiltonian for this system is expressed as:

\begin{align}
H_c = \sum_{i,j = 0}^3 u_{i+j}(t) \sigma_i^{(1)} \sigma_j^{(2)},
\end{align}

where $i \neq j \neq 0$. \(\sigma_i^{(1)}\) and \(\sigma_j^{(2)}\) represent the Pauli operators acting on the first and second particles, and the control coefficients \(u_{i+j}(t)\) represent the control field acting on the two qubits. This formulation facilitates a complete exploration of the two-qubit Hilbert space, with control operations applied either locally (i.e., when \( i \) or \( j = 0 \)) or globally (i.e., when \( i, j \neq 0 \)) to optimize the interaction between qubits. Figure \ref{fig:schemes} illustrates the schemes in which QOC is applied. Such coupling interactions are often adopted in Heisenberg spin model for quantum enhancement \cite{SKS:23:RMP,AAR:APB:24,AAH:24:PRA}.

\begin{figure*}[t]
\hspace*{-1cm}
  \begin{tikzpicture}
    \node[anchor=south west, inner sep=0] (image) at (0,0) 
      {\includegraphics[width=\textwidth]{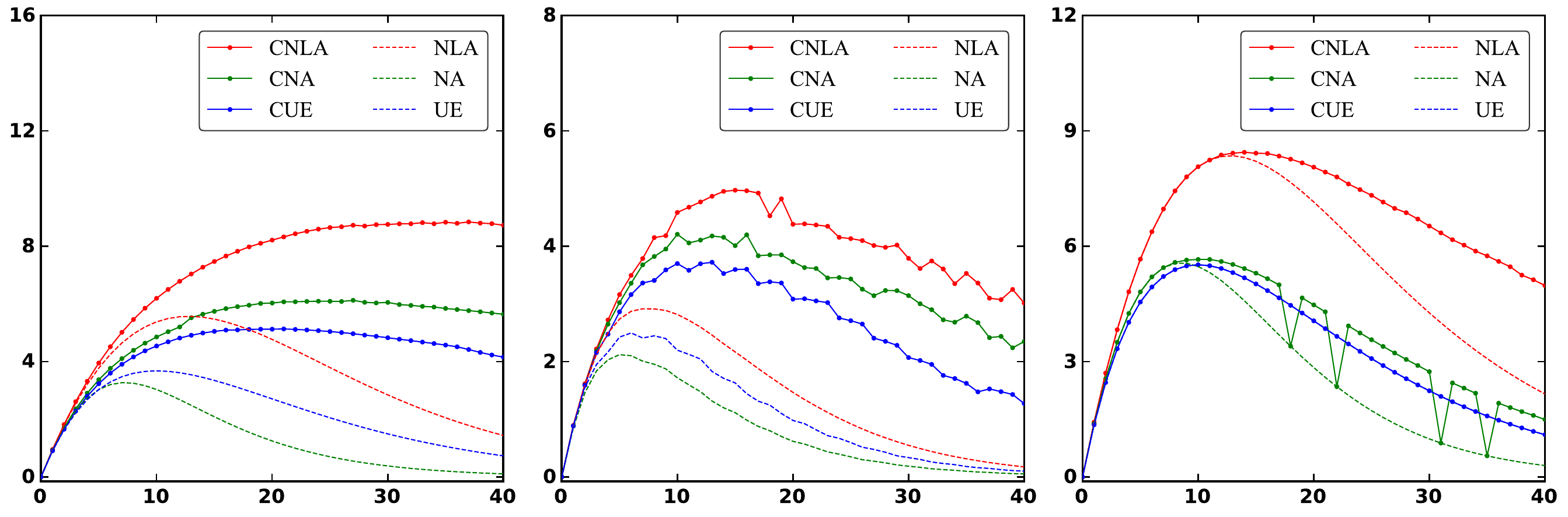}};
      
    
    \node at (3.09,-0.2) {$\boldsymbol{T}$}; 
    \node at (-0.19,3.173) {$\boldsymbol{\frac{\mathcal{F}_Q}{T}}$};
    \node at (0.97,5.32) {\textbf{(a)}}; 
    \node at (9.07,-0.2) {$\boldsymbol{T}$}; 
    \node at (7.01,5.32) {\textbf{(b)}}; 

    \node at (15.05,-0.2) {$\boldsymbol{T}$};
    \node at (13.05,5.32) {\textbf{(c)}}; 

  \end{tikzpicture}

\caption{\raggedright The performance of $\frac{\mathcal{F}_Q}{T}$ with increasing evolution time $T$ for \textbf{(a)} spontaneous emission, \textbf{(b)} generalized Pauli dephasing \textbf{(c)} Pauli-$XY$. Pauli-$XY$ dephasing severely limits the QOC scheme in the single-particle scheme with no apparent improvement in the normalized QFI with increasing $T$.}
\label{fig:QFI_time}
\end{figure*}
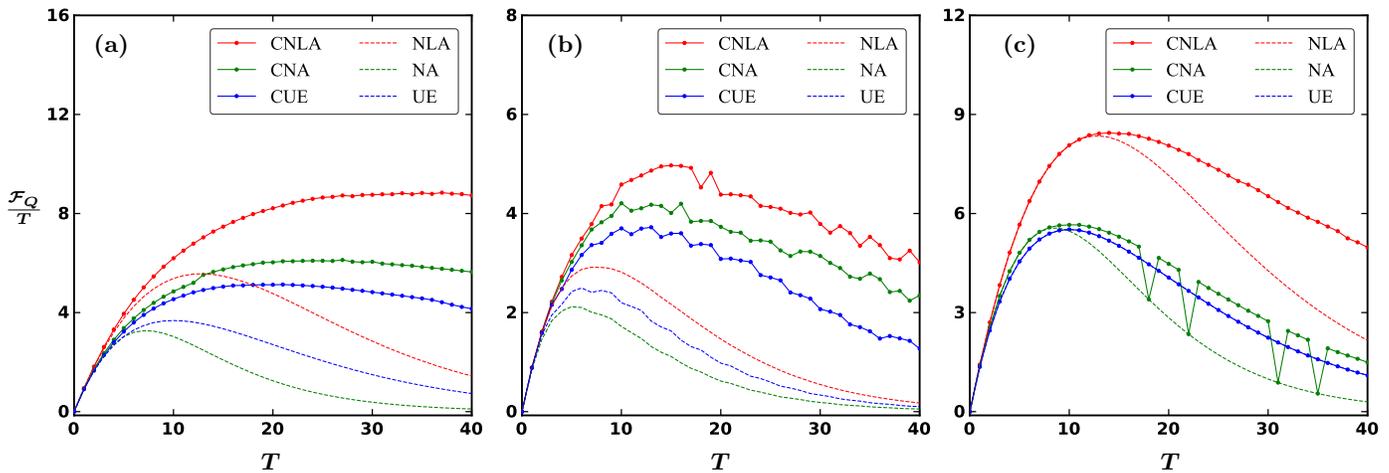

\section{Evaluation} \label{sec:app}
In this section, we perform and analyze numerical simulations for time-homogenous and time-inhomogeneous scenarios. We compare the performance of these protocols under both controlled and uncontrolled conditions using the normalized QFI, $\frac{\mathcal{F}_Q}{T}$, as a metric which is evaluated using \ref{eq: eq_3}.  For numerical simulations, the initial state of the single qubit is assumed to be $\ket{\psi} = \frac{\ket{0} + \ket{1}}{\sqrt{2}}$. In the entanglement-enhanced schemes, we employ the entangled state $\Phi = \frac{\ket{00} + \ket{11}}{\sqrt{2}}$. The learning rate is set to $l_r = 0.001$, and the time step for the control process is $\Delta t = 0.01$.

\subsection{Time Homogenous Markovian Evolution}

\subsubsection{Spontaneous Emission}
Spontaneous emission can be represented by the following variation to \ref{eq: lind_2}:

\begin{equation}
\begin{split}
    \dot{\rho}(t) = \iota[H,\rho] +\sum_{i=1}^k \gamma_{i} \big[ L_{-}^{(i)} \rho(t) L_{+}^{(i)} - \frac{1}{2}{L_{+}^{(i)} L_{-}^{(i)} ,\rho(t)} \big]
\end{split}
\end{equation}

\( L_{\pm}^{(i)} \) is the local operator acting on the \( i^{\text{th}} \) particle and \( L_{\pm} = \left( \sigma_1 \pm \iota \sigma_2 \right) \)/$2$ are the ladder operators. For the single qubit and noiseless ancilla schemes, we have \( k=1 \) and \( \gamma_1 = 0.1 \). The operator acts locally for the noiseless ancilla case, with local operations defined as \( L_{\pm}^{(1)} = L_{\pm} \otimes \mathcal{I} \) and \( L_{\pm}^{(2)} = \mathcal{I} \otimes L_{\pm} \) with \( \gamma_{2} = 0.05 \) for the noisy ancilla scheme.

In Fig. \ref{fig:QFI_time}(a), we monitor the behavior of the uncontrolled schemes  (NLA, NA, UE) and the controlled schemes (CNLA, CNA, CUE). As shown in the figure, we observe that the controlled schemes perform much better in all three cases than the uncontrolled schemes. The NLA scheme performs better than the other two uncontrolled schemes (NA and UE) and also outperforms some controlled schemes (CUE and CNA) for shorter evolution time. However, when the time increases beyond $T= 15$, we observe relative robustness to increasing noise in the NA scheme and the CUE scheme. In this case, we still observe a relatively improved performance of the CNA scheme compared to the CUE scheme. The behavior of the CNLA scheme is significantly better, and we observe a significant improvement in performance, and $\frac{\mathcal{F}_Q}{T}$ remains relatively stable toward the end of the graph, depicting the robustness that QOC can provide. Furthermore, regardless of a noisy ancilla, QOC provides better performance against the CUE scheme.

 \begin{figure*}[t]
 \hspace*{-0.8cm}
  \begin{tikzpicture}
    \node[anchor=south west, inner sep=0] (image) at (0,0) 
      {\includegraphics[width=\textwidth]{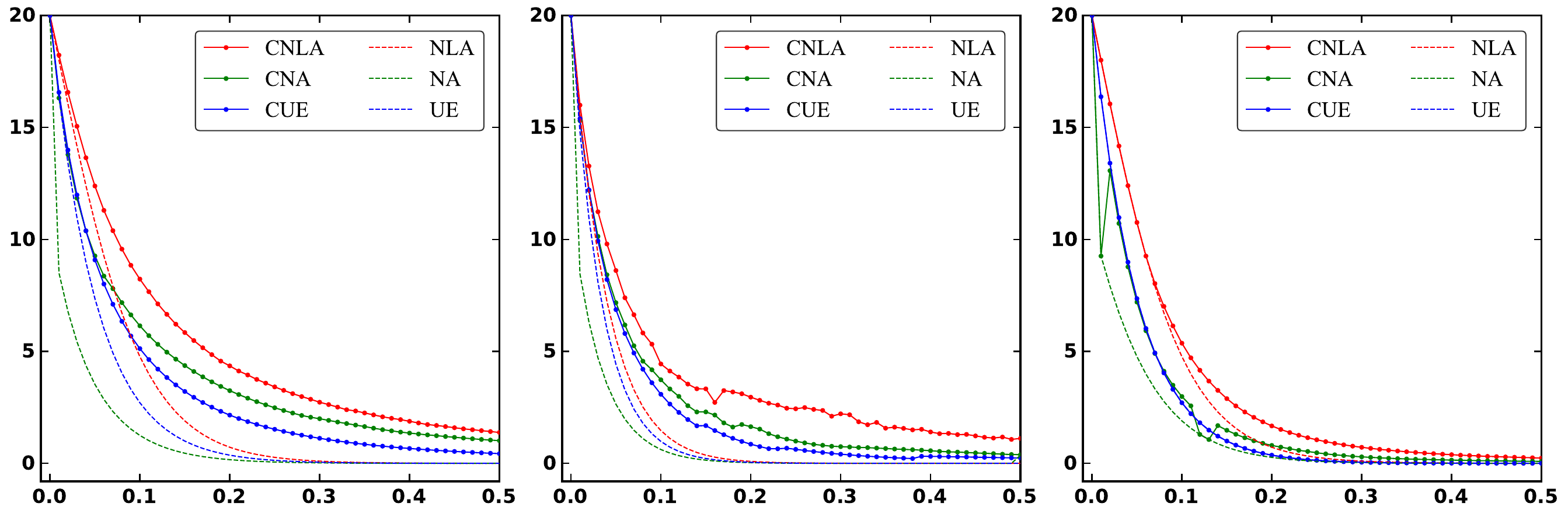}};
      
    \node at (3.00,-0.2) {$\boldsymbol{\gamma_1}$}; 
    \node at (-0.19,3.20) {$\boldsymbol{\frac{\mathcal{F}_Q}{T}}$};
    \node at (1.03,5.32) {\textbf{(a)}};
    \node at (9.13,-0.2) {$\boldsymbol{\gamma_1}$}; 
    \node at (7.04,5.32) {\textbf{(b)}}; 

    \node at (15.08,-0.2) {$\boldsymbol{\gamma_1}$};
    \node at (13.05,5.32) {\textbf{(c)}}; 

  \end{tikzpicture}
\caption{\raggedright The normalized QFI plotted against increasing $\gamma_1$ applied on the sensing particle. For the cases of noisy ancilla(CNA and NA), we keep $\gamma_2$ constant. The noise models depicted are: \textbf{(a)} spontaneous emission, \textbf{(b)} generalized Pauli dephasing, and \textbf{(c)} Pauli-XY. }
  \label{fig:Decay}
\end{figure*}
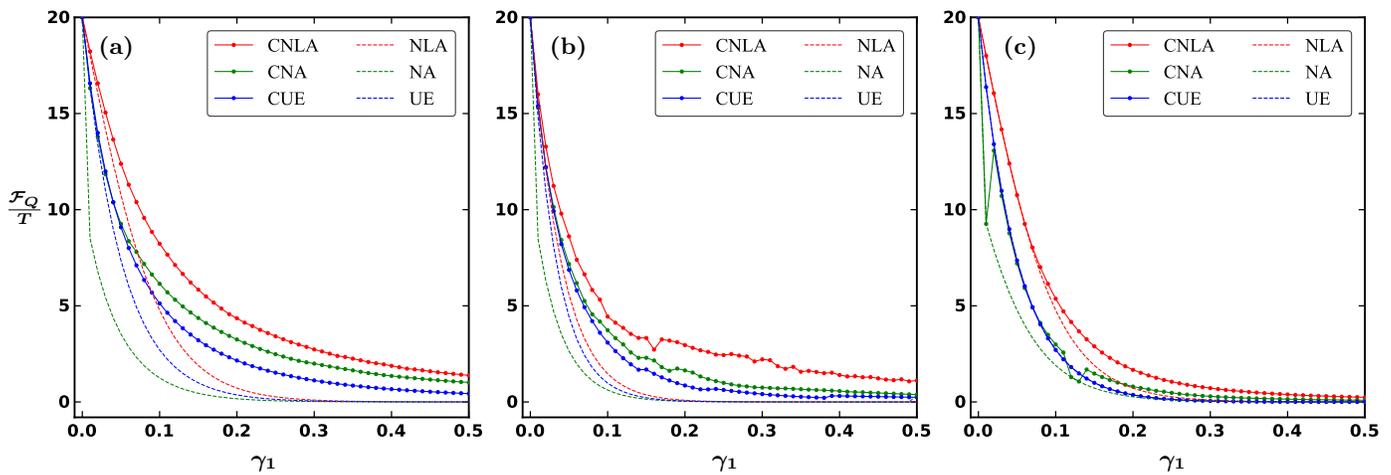

\subsubsection{ Generalized Pauli Dephasing}\label{sec:pauli}

Under dephasing dynamics, the master equation in Eq.~\ref{eq: lind_2} is modified as:

\begin{equation}
\dot{\rho}(t) = \iota [H(t), \rho(t)] + \sum_{i=1}^{k} \frac{\gamma_i}{2} \left[ L_{gp}^{(i)} \rho(t) {L_{gp}^{(i)}} - \rho(t) \right],
\end{equation}

\noindent where $L_{gp}^{(i)}$ represents the Lindblad operator acting locally on the $i^{\text{th}}$ particle. The Lindblad operator  $L_{gp}$ is defined as  $L_{gp} = n \cdot \sigma_n$, where $n$ is the unit vector given by  $n = (\cos(\theta)\sin(\phi), \sin(\theta)\sin(\phi), \cos(\theta))$, and  $\sigma_n = (\sigma_1, \sigma_2, \sigma_3)$ represents the Pauli matrices.

For both single-particle and bipartite schemes, the number of particles experiencing parameter-dependent noisy evolution, $k = 1$, and the decoherence rate is $\gamma_1 = 0.05$. In contrast, for the noisy ancilla scheme, $k = 2$, with decoherence rates $\gamma_1 = 0.1$ and $\gamma_2 = 0.05$. In this analysis, $\theta = \frac{\pi}{4}$ and $\phi = 0$ are used.

Fig. \ref{fig:QFI_time}(b) shows that the NLA and UE schemes outperform the NA scheme due to the additional decoherence experienced by the ancilla. This disparity becomes more pronounced as $T$ increases, particularly around $T=10$, after which the increase in noise diminishes the efficacy of the protocols. This occurs because, with larger $T$, the decoherence effects overshadow any improvement in $\frac{\mathcal{F}_Q}{T}$ provided by the parameter map.

When QOC is applied, we observe a significant improvement in $\frac{\mathcal{F}_Q}{T}$. The CNA scheme outperforms the CUE scheme and, as time progresses, counters decoherence much more efficiently than the UE scheme. The CNLA scheme shows superior performance compared to both, and, like the CNA scheme, offers substantial improvement over the UCE scheme. This indicates a robustness to increasing decoherence, helping to better sustain the advantage in $\frac{\mathcal{F}_Q}{T}$.

\subsubsection{Pauli-$XY$ Noise}

Pauli-$XY$ noise is crucial in state-of-the-art QM experiments, such as those involving NV centers \cite{SSK:17:Quantum,BCK:15:PRX} and is particularly challenging to address. The noise can be represented by :

\begin{equation}
\begin{split}
    \dot{\rho}(t) = \iota[H,\rho] +\sum_{i=1}^k \sum_{j=1}^2 \frac{\gamma_{i}}{2} p_j   \left[ \sigma_j^{(i)} \rho(t) \sigma_j^{(i)} - \frac{1}{2}\rho(t)\right]  
\end{split}
\end{equation}

\noindent where $\sigma_j^{(i)}$ denotes the local operation on the $i^{\text{th}}$ particle. Additionally, $p_1 = p$ and $p_2 = 1 - p$, with $p$ characterizing the degree of asymmetry. For both the single-particle and bipartite schemes with a noiseless ancilla, there is one particle undergoing parameter-dependent noisy evolution ($k = 1$) with a decoherence rate of $\gamma_1 = 0.1$. In contrast, for the noisy ancilla scheme, two particles undergo noisy evolution ($k = 2$) with rates $\gamma_1 = 0.1$ and $\gamma_2 = 0.05$. In Fig. \ref{fig:QFI_time}(c), the performance of different strategies against Pauli-$XY$ noise with $p=0.5$ is shown.

The UE scheme is severely limited by the noise, and interestingly, adding QOC provides no apparent advantage. However, for the entanglement-enhanced schemes, both the CNLA and CUE schemes outperform their uncontrolled counterparts. As time progresses, QOC provides additional robustness against noise for the CNLA scheme, resulting in significantly better overall performance compared to other strategies. A similar trend is observed for the CNA scheme. With increasing $T$; the robustness of the NA scheme improves through QOC. Again, despite the presence of a noisy ancilla, the performance still surpasses that of the CUE scheme.

\begin{figure*}[t]
\hspace*{-0.8cm}
  \begin{tikzpicture}
    \node[anchor=south west, inner sep=0] (image) at (0,0) 
      {\includegraphics[width=\textwidth]{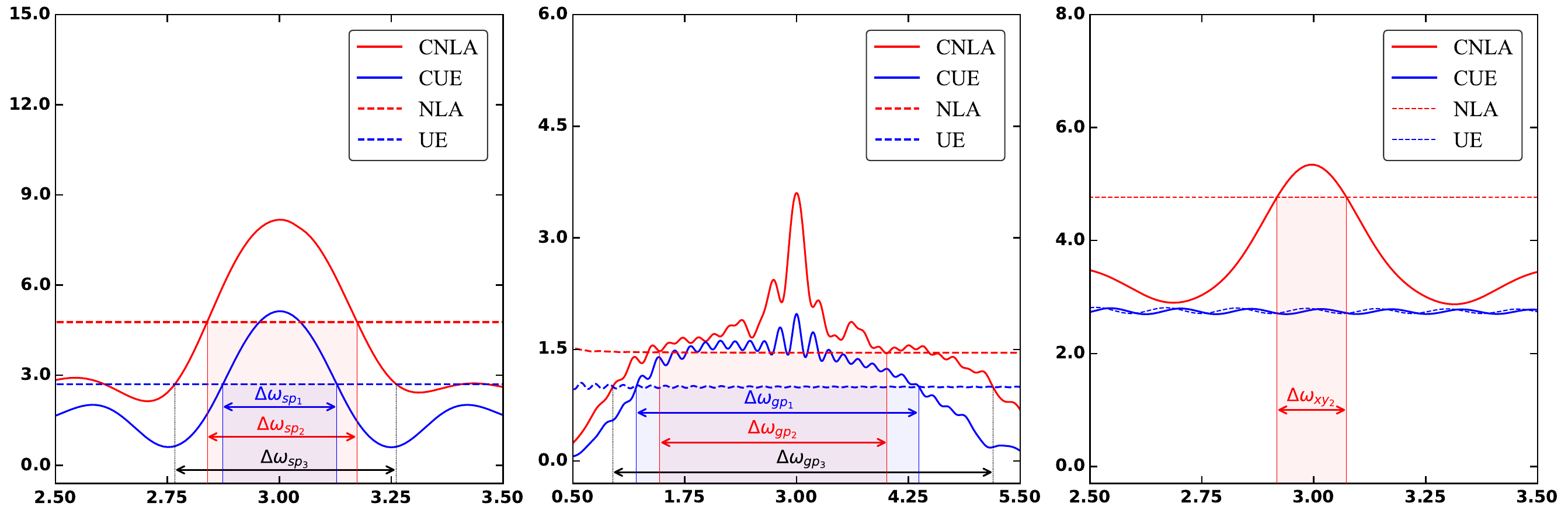}};
      
    \node at (3.135,-0.2) {$\boldsymbol{\omega_0}$}; 
    \node at (-0.055,3.217) {$\boldsymbol{\frac{\mathcal{F}_Q}{T}}$};
    \node at (1.15,5.25) {\textbf{(a)}}; 
    \node at (9.13,-0.2) {$\boldsymbol{\omega_0}$}; 
    \node at (7.145,5.25) {\textbf{(b)}}; 

    \node at (15.26,-0.2) {$\boldsymbol{\omega_0}$};
    \node at (13.14,5.25) {\textbf{(c)}}; 

  \end{tikzpicture}
  \caption{\raggedright The performance of the noiseless ancilla and unentangled scheme for $\omega_0$ variation. The dotted lines depict the uncontrolled versions of the schemes while solid lines depict the same schemes with QOC applied for noise types: \textbf{(a)} spontaneous emission, \textbf{(b)} generalized Pauli dephasing, and \textbf{(c)} Pauli-$XY$ noise for $T=20$.}
  \label{fig:generalizability}
\end{figure*}
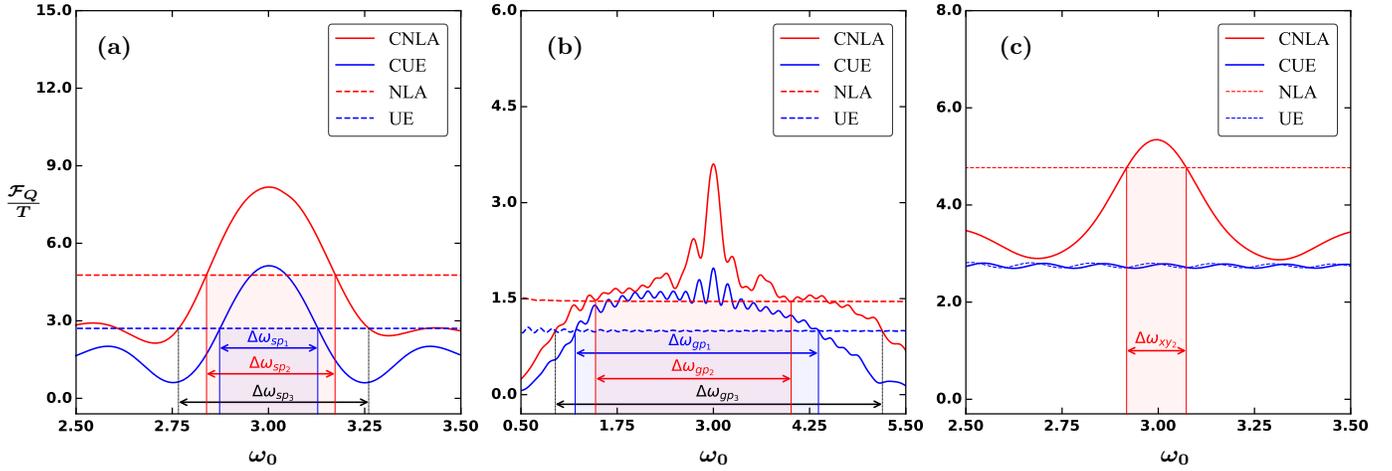

\subsubsection{Performance in High Noise Regime}

We now analyze how variations in the noise parameter affect the achievable QFI. For all three noise types, we vary $\gamma_1$ from $0$ to $0.5$ while keeping $\gamma_2$ fixed at 0.5.

As shown in Fig. \ref{fig:Decay}, QOC mitigates a sharp decline in QFI. We first examine the case of spontaneous emission, as depicted in Fig. \ref{fig:Decay}(a). Compared to the Pauli dephasing channel in Fig. \ref{fig:Decay}(b), all schemes display greater robustness against changes in $\gamma_1$, with a more pronounced variation in the decline of $\frac{F_Q}{T}$. At $\gamma_1 = 0$, all schemes begin with a high $\frac{F_Q}{T}$ value around $20$. As the noise parameter increases, controlled schemes consistently outperform uncontrolled ones, with the uncontrolled sequential scheme exhibiting the steepest decline, as expected.

The controlled noisy ancilla scheme significantly improves upon the sequential scheme but still slightly underperforms relative to the uncontrolled noiseless ancilla case. Both probe types lose coherence at similar values of $\gamma$. In the low-noise regime, the controlled noisy ancilla and controlled sequential schemes perform comparably. However, in the high-noise regime, the noisy ancilla scheme outperforms the sequential scheme. As expected, the controlled noiseless ancilla scheme provides the best performance, showing the slowest rate of decline as $\gamma$ increases.

In the specific case of Pauli-$XY$ noise, shown in Fig. \ref{fig:Decay}(c), the uncontrolled schemes suffer substantial degradation as the noise parameter increases. In contrast, controlled schemes, particularly those that use a controlled ancilla, exhibit improved robustness to noise. Among them, the noiseless ancilla scheme delivers the best performance, clearly outperforming the other strategies. However, the CUE scheme offers no advantage over the UE scheme, while the CNA scheme surpasses the CUE scheme in the high-noise regime.

\subsubsection{Scheme Robustness to Parameter Variation}

In QM, optimal control strategies are typically tailored to a specific parameter of interest; in this case, \(\omega = 3\). Although these optimized controls are effective for the target parameter, their performance may degrade when the parameter deviates from this value within the specified noise model.

To address this limitation, it is essential to develop schemes that maintain an advantage over uncontrolled approaches across a broader spectrum of parameter values. In Fig. \ref{fig:generalizability}, we compare the performance of models utilizing entangled and non-entangled schemes, excluding the noisy ancilla case for simplicity. We derive optimal controls and vary the parameter values to evaluate the robustness of the QM schemes.

The CNLA scheme demonstrates a significant advantage over the UE scheme across a wider range of parameter values for the three noise models, as shown in Fig. \ref{fig:generalizability}, which corresponds to \(\omega_{sp_3}\), \(\omega_{gp_3}\), and \(\omega_{xy_3}\), respectively. For spontaneous emission in Fig. \ref{fig:generalizability}(a) and generalized Pauli dephasing in Fig. \ref{fig:generalizability}(b), the improvement of the CUE scheme over the UE scheme, represented by \(\omega_{sp_1}\), \(\omega_{gp_1}\), and \(\omega_{xy_1}\), is notably less than that provided by the CNLA scheme. Additionally, we depict the range of \(\omega\) values where the CNLA scheme outperforms the NLA scheme, indicated by \(\omega_{sp_2}\), \(\omega_{gp_2}\), and \(\omega_{xy_2}\). This advantage is particularly pronounced in the case of spontaneous emission, while the other noise models do not exhibit a clear superiority of the CNLA scheme over the CUE scheme.

For the Pauli-$XY$ noise represented in Fig. \ref{fig:generalizability}(c), we omit \(\omega_{xy_1}\) and \(\omega_{xy_3}\), as the \(\frac{\mathcal{F}_Q}{T}\) for the CNLA schemes does not fall below that of the UE or CUE schemes in this specific case.

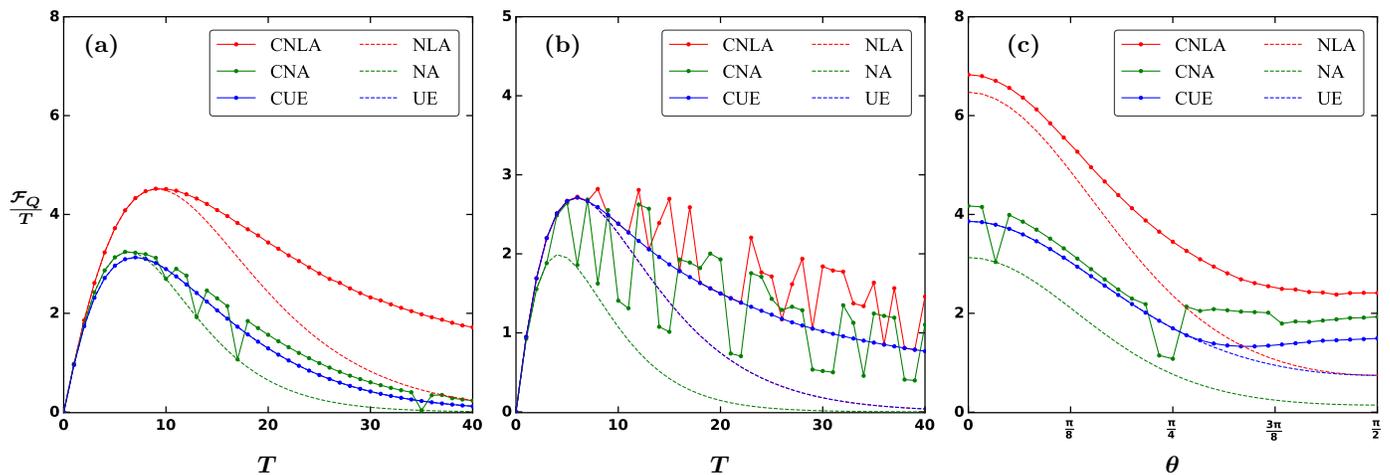
\begin{figure*}[t]

\hspace*{-0.8cm}
\begin{tikzpicture}
    \node[anchor=south west, inner sep=0] (image) at (0,0) 
      {\includegraphics[width=\textwidth]{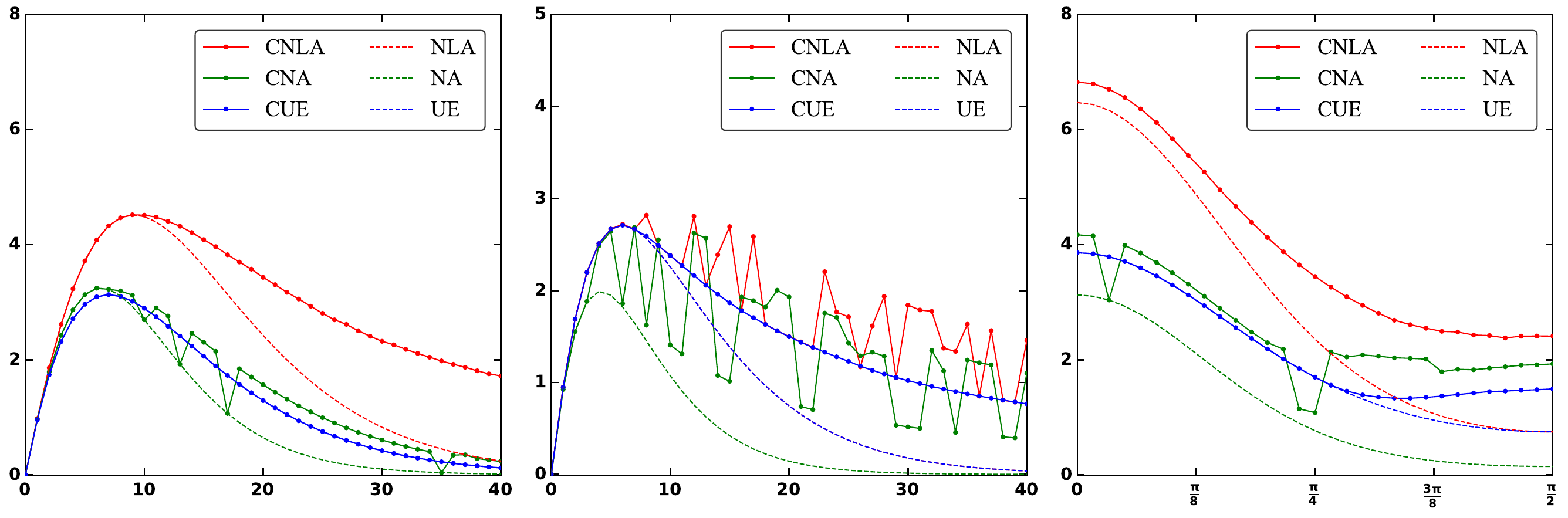}};
      
    \node at (3.00,-0.2) {$\boldsymbol{T}$}; 
    \node at (-0.205,3.217) {$\boldsymbol{\frac{\mathcal{F}_Q}{T}}$};
    \node at (0.8,5.37) {\textbf{(a)}}; 
    \node at (9.02,-0.2) {$\boldsymbol{T}$}; 
    \node at (6.93,5.37) {\textbf{(b)}}; 
    \node at (15.04,-0.2) {$\boldsymbol{\theta}$};
    \node at (13.06,5.37) {\textbf{(c)}}; 

\end{tikzpicture}

\caption{\raggedright \textbf{(a)} shows the $\frac{\mathcal{F}_Q}{T}$ performance with time inhomogeneous dissipation, \textbf{(b)} shows the QFI performance in the case of time inhomogeneous dephasing, and \textbf{(c)} shows the transition between dephasing and dissipative dynamics for $T=20$.}
\label{fig:time_inhomogenous}
\end{figure*}

\subsection{Time-Inhomogeneous Markovian Evolution}
In this section, we analyze the case of noisy frequency estimation typically encountered in atomic spectroscopy experiments. We focus on the decoherence modeled by single qubit dynamics with the spin-boson model given as:

\begin{align}
H =& \frac{\omega_0 \sigma_z}{2} + \sum_n \omega_n a_n^{\dagger} a_n + 
\left(cos(\theta)\frac{\sigma_x}{2} +sin(\theta) \frac{\sigma_z}{2}\right)  \nonumber \\ 
& \otimes \sum_n (g_n a_n +g_n^* a_n^\dagger) 
\end{align}

\noindent where the first term is the parameter encoding Hamiltonian $H_{\omega_0}$, the second is the bath Hamiltonian $H_B$ and the third is the interaction Hamiltonian $H_I$. $a_n^\dagger$ and $a_n$ are the creation and annihilation operators for the $n$-th mode of the bath with frequency $\omega_n$. $g_n$ represents the coupling strength between the bath and the system, and the parameter $\theta$ represents the coupling angle. We will utilize the dynamics obtained in the high-temperature regime while assuming the rotating wave approximation and the spectral density of the bath to be Ohmic. The resulting dynamics are PC and are given by \cite{HSK:18:NJP}:

\begin{align}
\frac{d\rho}{dt} = & -i[H(t), \rho] +  \nonumber \\ & \sum_{i,j} \gamma_i(t) d_j \left( L^{(i)}_j \rho {L_j^{(i)}}^\dagger - \frac{1}{2} {L_j^{(i)}}^\dagger L_j^{(i)}, \rho  \right),
\end{align}

\noindent where $i = \{1,2\}$ and $j = \{\pm,z\}$ with $d_{\pm} = \cos\left(\theta\right)^2$ and $d_z = \sin\left(\theta\right)^2$. $i$ indicates the number of particles that undergo noisy evolution. The dynamics remain completely positive (CP) and, due to the positivity of \(\gamma(t)\) and \(d_j\), are also CP-divisible. The expression for the time-dependent decoherence rate is:

\begin{equation} \label{eq:gamma}
\gamma_i(t) = \eta_i \arctan(w_c T),
\end{equation}

\noindent where $\omega_c$ is the cut-off frequency for the bath modes and $\eta = \frac{\lambda}{\beta}$  with $\lambda$ being the system-environment interaction strength and $\beta$ being the inverse temperature. For simplicity, we assume $\omega_c=1$. We will vary $\theta$ to evaluate the performance of optimal control in different $d_j$. Specifically, $\theta = 0$ corresponds to fully dissipative dynamics, while $ \theta = \frac{\pi}{2} $ corresponds to fully dephasing dynamics.

\subsubsection{Dissipative Dynamics}

In the context of dissipative dynamics, as shown in Fig. \ref{fig:time_inhomogenous}(a), we vary $T$ to examine how the protocol performs over time, under constantly evolving decoherence. Under the assumption of a constant $\gamma$, previous results indicated that the CNLA scheme demonstrated significant robustness against spontaneous emission, particularly at higher values of $T$. The CNA and CUE schemes also exhibited strong performance.

However, in this scenario, we incorporate the effects of spontaneous absorption, which simultaneously impacts the system. Furthermore, as $T$ increases, the protocol faces rising values of $\gamma(t)$ from Eq. \ref{eq:gamma}, which introduces further limitations. For this analysis, we take $\eta_1 = 0.05$. Under these conditions, the performance is not as ideal as before. However, the CNLA scheme consistently outperforms the others. Although its robustness to noise decreases as $T$ increases, it still maintains superior performance compared to the other protocols.

Next, we consider the noisy ancilla case, where we apply a similar time-dependent decay to the ancillary qubit but with a lower decoherence rate of $\eta_2 = 0.025$, while keeping the same value of $\eta_1$ as in the NLA case. Here, the noisy ancilla scheme continues to provide greater noise robustness than the UE scheme. Despite some sudden dips in performance, the overall trend remains relatively stable, with the convergence to $F_Q \rightarrow 0$ occurring more slowly than in the single-particle case.

For the single-particle scheme, we observe that regardless of the applied control field, there is no improvement in performance for large $T$. This result highlights a key finding: dynamical controls on a single qubit may fail to adapt effectively in a fully dissipative environment, underscoring the necessity of an ancillary qubit to enhance noise robustness in scenarios dominated by dissipative noise.

\subsubsection{Dephasing Dynamics}

For the time-homogeneous case, we observed that the protocols exhibited relatively stable behavior with respect to $T$ when subjected to a generalized Pauli dephasing. However, transitioning to complete dephasing dynamics, as shown in Fig. \ref{fig:time_inhomogenous}(b) with $\theta = \frac{\pi}{2}$, presents significant challenges. Noise that aligns with the parameter-encoding Hamiltonian can greatly diminish the accuracy of parameter estimation. In the case of the CUE scheme, the control protocol performs reasonably well, with control mechanisms enhancing robustness as noise levels increase. For this scenario, we assume $\eta_1 = 0.03$ and $\eta_2 = 0.015$.

For the noisy ancilla scenario, the NA scheme faces severe performance limitations. Although the introduction of QOC improves the scheme’s behavior, the normalized QFI displays noisy fluctuations, suggesting that pure dephasing significantly restricts the controls' ability to converge to an optimal value for a given $T$. The $\frac{F_Q}{T}$ values oscillate around those of the CUE scheme, but no clear advantage is observed.

In contrast, the CNLA scheme outperforms the other strategies. Although convergence to the optimal value for a given $T$ remains noisy, the CNLA curve consistently oscillates above that of the single-particle scheme, indicating that the CNLA scheme achieves slightly superior performance across the examined time steps.

\subsubsection{\texorpdfstring{\(\theta\) Variation}{Theta Variation}}

We now explore the effect of varying $\theta$ on the controlled protocols for the entanglement-enhanced and single-particle cases, as shown in Fig. \ref{fig:time_inhomogenous}(c). This analysis focuses on how performance changes as $\theta$ transitions from $0$ (Dissipative Dynamics) to $\frac{\pi}{2}$ (Dephasing Dynamics). 

Initially, the entanglement-enhanced schemes demonstrate better performance than the single-particle control case. However, as $\theta$ increases, the performance gap between the CNLA and CUE schemes narrow, while the CNA scheme throughout gives a slightly better performance than the CUE scheme. When $\theta$ approaches $\frac{\pi}{2}$, the advantage of an active ancilla diminishes, and the single-particle case begins to perform better as dephasing dynamics dominate. Nevertheless, the noiseless ancilla scheme continues to provide superior performance across most values of $\theta$, which was to be expected.

\section{Conclusion and Outlook}\label{sec:dis}
In this paper, we demonstrate the effectiveness of our entanglement-enhanced strategy for quantum parameter estimation in comparison to the conventionally employed CUE scheme. Our results show that the entanglement-enhanced approach achieves significantly higher QFI, enabling improved precision even under heightened noise intensity and extended evolution times, including scenarios involving a noisy ancilla. Furthermore, this advantage extends over a wider range of parameter values, enhancing the robustness of the scheme against parameter fluctuations.

Our findings show that the presence of an ancilla enhances the effectiveness of QOC, providing greater adaptability to more challenging noise environments. This improvement is particularly evident in two scenarios: (i) under time-homogeneous Pauli-$XY$ noise and (ii) in time-inhomogeneous dissipative dynamics. In both cases, the CUE scheme offers no performance improvements over its uncontrolled counterpart, whereas the entanglement-enhanced strategy delivers a clear advantage.

Our findings suggest that using the entanglement-enhanced approach for QOC-enabled QM provides a more robust scheme to tackle general Markovian dynamics. Extending this study to multiqubit entangled systems will potentially provide an even greater improvement, which necessitates the study of protocols that can employ QOC for entangled quantum sensing systems. This is a challenging task due to the exponential growth of optimization parameters with the increasing size of the Hilbert space \cite{KBCD:22:EPJ,LJSK:24:JPC}. Recently, algorithms have been proposed to optimally control larger quantum systems \cite{WOKW:23:QS,RWD:24:CPC}, and it will be interesting to see how they are used for multiqubit QM. Future protocols must be carefully designed to balance the trade-off between computational complexity and metrological performance of the scheme in the presence of general noise models.

\section*{Disclosures}
The authors declare that they have no known competing financial interests.

\section*{Data availability}
The code and datasets generated and analyzed during this study are available from the corresponding author upon reasonable request.

\bibliographystyle{apsrev4-2}

\bibliography{manuscript} 

\begin{thebibliography}{41}%
\makeatletter
\providecommand \@ifxundefined [1]{%
 \@ifx{#1\undefined}
}%
\providecommand \@ifnum [1]{%
 \ifnum #1\expandafter \@firstoftwo
 \else \expandafter \@secondoftwo
 \fi
}%
\providecommand \@ifx [1]{%
 \ifx #1\expandafter \@firstoftwo
 \else \expandafter \@secondoftwo
 \fi
}%
\providecommand \natexlab [1]{#1}%
\providecommand \enquote  [1]{``#1''}%
\providecommand \bibnamefont  [1]{#1}%
\providecommand \bibfnamefont [1]{#1}%
\providecommand \citenamefont [1]{#1}%
\providecommand \href@noop [0]{\@secondoftwo}%
\providecommand \href [0]{\begingroup \@sanitize@url \@href}%
\providecommand \@href[1]{\@@startlink{#1}\@@href}%
\providecommand \@@href[1]{\endgroup#1\@@endlink}%
\providecommand \@sanitize@url [0]{\catcode `\\12\catcode `\$12\catcode `\&12\catcode `\#12\catcode `\^12\catcode `\_12\catcode `\%12\relax}%
\providecommand \@@startlink[1]{}%
\providecommand \@@endlink[0]{}%
\providecommand \url  [0]{\begingroup\@sanitize@url \@url }%
\providecommand \@url [1]{\endgroup\@href {#1}{\urlprefix }}%
\providecommand \urlprefix  [0]{URL }%
\providecommand \Eprint [0]{\href }%
\providecommand \doibase [0]{https://doi.org/}%
\providecommand \selectlanguage [0]{\@gobble}%
\providecommand \bibinfo  [0]{\@secondoftwo}%
\providecommand \bibfield  [0]{\@secondoftwo}%
\providecommand \translation [1]{[#1]}%
\providecommand \BibitemOpen [0]{}%
\providecommand \bibitemStop [0]{}%
\providecommand \bibitemNoStop [0]{.\EOS\space}%
\providecommand \EOS [0]{\spacefactor3000\relax}%
\providecommand \BibitemShut  [1]{\csname bibitem#1\endcsname}%
\let\auto@bib@innerbib\@empty
\bibitem [{\citenamefont {Giovannetti}\ \emph {et~al.}(2004)\citenamefont {Giovannetti}, \citenamefont {Lloyd},\ and\ \citenamefont {Maccone}}]{GLM:04:Science}%
  \BibitemOpen
  \bibfield  {author} {\bibinfo {author} {\bibfnamefont {V.}~\bibnamefont {Giovannetti}}, \bibinfo {author} {\bibfnamefont {S.}~\bibnamefont {Lloyd}},\ and\ \bibinfo {author} {\bibfnamefont {L.}~\bibnamefont {Maccone}},\ }\href@noop {} {\bibfield  {journal} {\bibinfo  {journal} {Science}\ }\textbf {\bibinfo {volume} {306}},\ \bibinfo {pages} {1330} (\bibinfo {year} {2004})}\BibitemShut {NoStop}%
\bibitem [{\citenamefont {Giovannetti}\ \emph {et~al.}(2006)\citenamefont {Giovannetti}, \citenamefont {Lloyd},\ and\ \citenamefont {Maccone}}]{GLM:06:PRL}%
  \BibitemOpen
  \bibfield  {author} {\bibinfo {author} {\bibfnamefont {V.}~\bibnamefont {Giovannetti}}, \bibinfo {author} {\bibfnamefont {S.}~\bibnamefont {Lloyd}},\ and\ \bibinfo {author} {\bibfnamefont {L.}~\bibnamefont {Maccone}},\ }\href@noop {} {\bibfield  {journal} {\bibinfo  {journal} {Phys. Rev. Lett.}\ }\textbf {\bibinfo {volume} {96}},\ \bibinfo {pages} {010401} (\bibinfo {year} {2006})}\BibitemShut {NoStop}%
\bibitem [{\citenamefont {Giovannetti}\ \emph {et~al.}(2011)\citenamefont {Giovannetti}, \citenamefont {Lloyd},\ and\ \citenamefont {Maccone}}]{GLM:11:Nature}%
  \BibitemOpen
  \bibfield  {author} {\bibinfo {author} {\bibfnamefont {V.}~\bibnamefont {Giovannetti}}, \bibinfo {author} {\bibfnamefont {S.}~\bibnamefont {Lloyd}},\ and\ \bibinfo {author} {\bibfnamefont {L.}~\bibnamefont {Maccone}},\ }\href@noop {} {\bibfield  {journal} {\bibinfo  {journal} {Nature photonics}\ }\textbf {\bibinfo {volume} {5}},\ \bibinfo {pages} {222} (\bibinfo {year} {2011})}\BibitemShut {NoStop}%
\bibitem [{\citenamefont {T{\'o}th}\ and\ \citenamefont {Apellaniz}(2014)}]{TA:14:JPA}%
  \BibitemOpen
  \bibfield  {author} {\bibinfo {author} {\bibfnamefont {G.}~\bibnamefont {T{\'o}th}}\ and\ \bibinfo {author} {\bibfnamefont {I.}~\bibnamefont {Apellaniz}},\ }\href@noop {} {\bibfield  {journal} {\bibinfo  {journal} {Journal of Physics A: Mathematical and Theoretical}\ }\textbf {\bibinfo {volume} {47}},\ \bibinfo {pages} {424006} (\bibinfo {year} {2014})}\BibitemShut {NoStop}%
\bibitem [{\citenamefont {Pezze}\ \emph {et~al.}(2018)\citenamefont {Pezze}, \citenamefont {Smerzi}, \citenamefont {Oberthaler}, \citenamefont {Schmied},\ and\ \citenamefont {Treutlein}}]{LSOS:20:Quantum}%
  \BibitemOpen
  \bibfield  {author} {\bibinfo {author} {\bibfnamefont {L.}~\bibnamefont {Pezze}}, \bibinfo {author} {\bibfnamefont {A.}~\bibnamefont {Smerzi}}, \bibinfo {author} {\bibfnamefont {M.~K.}\ \bibnamefont {Oberthaler}}, \bibinfo {author} {\bibfnamefont {R.}~\bibnamefont {Schmied}},\ and\ \bibinfo {author} {\bibfnamefont {P.}~\bibnamefont {Treutlein}},\ }\href@noop {} {\bibfield  {journal} {\bibinfo  {journal} {Reviews of Modern Physics}\ }\textbf {\bibinfo {volume} {90}},\ \bibinfo {pages} {035005} (\bibinfo {year} {2018})}\BibitemShut {NoStop}%
\bibitem [{\citenamefont {Kołodyński}\ and\ \citenamefont {Demkowicz-Dobrzański}(2013)}]{KD:13:NJP}%
  \BibitemOpen
  \bibfield  {author} {\bibinfo {author} {\bibfnamefont {J.}~\bibnamefont {Kołodyński}}\ and\ \bibinfo {author} {\bibfnamefont {R.}~\bibnamefont {Demkowicz-Dobrzański}},\ }\href@noop {} {\bibfield  {journal} {\bibinfo  {journal} {New Journal of Physics}\ }\textbf {\bibinfo {volume} {15}},\ \bibinfo {pages} {073043} (\bibinfo {year} {2013})}\BibitemShut {NoStop}%
\bibitem [{\citenamefont {Demkowicz-Dobrza\ifmmode~\acute{n}\else \'{n}\fi{}ski}\ and\ \citenamefont {Maccone}(2014)}]{DM:14:PRL}%
  \BibitemOpen
  \bibfield  {author} {\bibinfo {author} {\bibfnamefont {R.}~\bibnamefont {Demkowicz-Dobrza\ifmmode~\acute{n}\else \'{n}\fi{}ski}}\ and\ \bibinfo {author} {\bibfnamefont {L.}~\bibnamefont {Maccone}},\ }\href@noop {} {\bibfield  {journal} {\bibinfo  {journal} {Phys. Rev. Lett.}\ }\textbf {\bibinfo {volume} {113}},\ \bibinfo {pages} {250801} (\bibinfo {year} {2014})}\BibitemShut {NoStop}%
\bibitem [{\citenamefont {Sekatski}\ \emph {et~al.}(2017)\citenamefont {Sekatski}, \citenamefont {Skotiniotis}, \citenamefont {Kołodyński},\ and\ \citenamefont {Dür}}]{SSK:17:Quantum}%
  \BibitemOpen
  \bibfield  {author} {\bibinfo {author} {\bibfnamefont {P.}~\bibnamefont {Sekatski}}, \bibinfo {author} {\bibfnamefont {M.}~\bibnamefont {Skotiniotis}}, \bibinfo {author} {\bibfnamefont {J.}~\bibnamefont {Kołodyński}},\ and\ \bibinfo {author} {\bibfnamefont {W.}~\bibnamefont {Dür}},\ }\href@noop {} {\bibfield  {journal} {\bibinfo  {journal} {Quantum}\ }\textbf {\bibinfo {volume} {1}},\ \bibinfo {pages} {27} (\bibinfo {year} {2017})}\BibitemShut {NoStop}%
\bibitem [{\citenamefont {Zhou}\ \emph {et~al.}(2018)\citenamefont {Zhou}, \citenamefont {Zhang}, \citenamefont {Preskill},\ and\ \citenamefont {Jiang}}]{ZZP:18:Nature}%
  \BibitemOpen
  \bibfield  {author} {\bibinfo {author} {\bibfnamefont {S.}~\bibnamefont {Zhou}}, \bibinfo {author} {\bibfnamefont {M.}~\bibnamefont {Zhang}}, \bibinfo {author} {\bibfnamefont {J.}~\bibnamefont {Preskill}},\ and\ \bibinfo {author} {\bibfnamefont {L.}~\bibnamefont {Jiang}},\ }\href@noop {} {\bibfield  {journal} {\bibinfo  {journal} {Nature communications}\ }\textbf {\bibinfo {volume} {9}},\ \bibinfo {pages} {78} (\bibinfo {year} {2018})}\BibitemShut {NoStop}%
\bibitem [{\citenamefont {Kurdziałek}\ \emph {et~al.}(2023)\citenamefont {Kurdziałek}, \citenamefont {Górecki}, \citenamefont {Albarelli},\ and\ \citenamefont {Demkowicz-Dobrzański}}]{KGAD:23:PRL}%
  \BibitemOpen
  \bibfield  {author} {\bibinfo {author} {\bibfnamefont {S.}~\bibnamefont {Kurdziałek}}, \bibinfo {author} {\bibfnamefont {W.}~\bibnamefont {Górecki}}, \bibinfo {author} {\bibfnamefont {F.}~\bibnamefont {Albarelli}},\ and\ \bibinfo {author} {\bibfnamefont {R.}~\bibnamefont {Demkowicz-Dobrzański}},\ }\href@noop {} {\bibfield  {journal} {\bibinfo  {journal} {Physical Review Letters}\ }\textbf {\bibinfo {volume} {131}},\ \bibinfo {pages} {090801} (\bibinfo {year} {2023})}\BibitemShut {NoStop}%
\bibitem [{\citenamefont {Chaves}\ \emph {et~al.}(2013)\citenamefont {Chaves}, \citenamefont {Brask}, \citenamefont {Markiewicz}, \citenamefont {Kołodyński},\ and\ \citenamefont {Acín}}]{CBM:13:PRL}%
  \BibitemOpen
  \bibfield  {author} {\bibinfo {author} {\bibfnamefont {R.}~\bibnamefont {Chaves}}, \bibinfo {author} {\bibfnamefont {J.~B.}\ \bibnamefont {Brask}}, \bibinfo {author} {\bibfnamefont {M.}~\bibnamefont {Markiewicz}}, \bibinfo {author} {\bibfnamefont {J.}~\bibnamefont {Kołodyński}},\ and\ \bibinfo {author} {\bibfnamefont {A.}~\bibnamefont {Acín}},\ }\href@noop {} {\bibfield  {journal} {\bibinfo  {journal} {Phys. Rev. Lett.}\ }\textbf {\bibinfo {volume} {111}},\ \bibinfo {pages} {120401} (\bibinfo {year} {2013})}\BibitemShut {NoStop}%
\bibitem [{\citenamefont {Albarelli}\ \emph {et~al.}(2018)\citenamefont {Albarelli}, \citenamefont {Rossi}, \citenamefont {Tamascelli},\ and\ \citenamefont {Genoni}}]{ARTG:18:Quantum}%
  \BibitemOpen
  \bibfield  {author} {\bibinfo {author} {\bibfnamefont {F.}~\bibnamefont {Albarelli}}, \bibinfo {author} {\bibfnamefont {M.~A.~C.}\ \bibnamefont {Rossi}}, \bibinfo {author} {\bibfnamefont {D.}~\bibnamefont {Tamascelli}},\ and\ \bibinfo {author} {\bibfnamefont {M.~G.}\ \bibnamefont {Genoni}},\ }\href@noop {} {\bibfield  {journal} {\bibinfo  {journal} {{Quantum}}\ }\textbf {\bibinfo {volume} {2}},\ \bibinfo {pages} {110} (\bibinfo {year} {2018})}\BibitemShut {NoStop}%
\bibitem [{\citenamefont {Ilias}\ \emph {et~al.}(2022)\citenamefont {Ilias}, \citenamefont {Yang}, \citenamefont {Huelga},\ and\ \citenamefont {Plenio}}]{IYHP:22:PRX}%
  \BibitemOpen
  \bibfield  {author} {\bibinfo {author} {\bibfnamefont {T.}~\bibnamefont {Ilias}}, \bibinfo {author} {\bibfnamefont {D.}~\bibnamefont {Yang}}, \bibinfo {author} {\bibfnamefont {S.~F.}\ \bibnamefont {Huelga}},\ and\ \bibinfo {author} {\bibfnamefont {M.~B.}\ \bibnamefont {Plenio}},\ }\href@noop {} {\bibfield  {journal} {\bibinfo  {journal} {PRX Quantum}\ }\textbf {\bibinfo {volume} {3}},\ \bibinfo {pages} {010354} (\bibinfo {year} {2022})}\BibitemShut {NoStop}%
\bibitem [{\citenamefont {Chin}\ \emph {et~al.}(2012)\citenamefont {Chin}, \citenamefont {Huelga},\ and\ \citenamefont {Plenio}}]{CHP:12:PRL}%
  \BibitemOpen
  \bibfield  {author} {\bibinfo {author} {\bibfnamefont {A.~W.}\ \bibnamefont {Chin}}, \bibinfo {author} {\bibfnamefont {S.~F.}\ \bibnamefont {Huelga}},\ and\ \bibinfo {author} {\bibfnamefont {M.~B.}\ \bibnamefont {Plenio}},\ }\href@noop {} {\bibfield  {journal} {\bibinfo  {journal} {Physical review letters}\ }\textbf {\bibinfo {volume} {109}},\ \bibinfo {pages} {233601} (\bibinfo {year} {2012})}\BibitemShut {NoStop}%
\bibitem [{\citenamefont {Altherr}\ and\ \citenamefont {Yang}(2021)}]{AY:21:PRL}%
  \BibitemOpen
  \bibfield  {author} {\bibinfo {author} {\bibfnamefont {A.}~\bibnamefont {Altherr}}\ and\ \bibinfo {author} {\bibfnamefont {Y.}~\bibnamefont {Yang}},\ }\href@noop {} {\bibfield  {journal} {\bibinfo  {journal} {Phys. Rev. Lett.}\ }\textbf {\bibinfo {volume} {127}},\ \bibinfo {pages} {060501} (\bibinfo {year} {2021})}\BibitemShut {NoStop}%
\bibitem [{\citenamefont {Yang}\ \emph {et~al.}(2024)\citenamefont {Yang}, \citenamefont {Long}, \citenamefont {Liu}, \citenamefont {Tang}, \citenamefont {Zhai}, \citenamefont {Nie}, \citenamefont {Xin}, \citenamefont {Li},\ and\ \citenamefont {Lu}}]{YLLT:24:Nature}%
  \BibitemOpen
  \bibfield  {author} {\bibinfo {author} {\bibfnamefont {X.}~\bibnamefont {Yang}}, \bibinfo {author} {\bibfnamefont {X.}~\bibnamefont {Long}}, \bibinfo {author} {\bibfnamefont {R.}~\bibnamefont {Liu}}, \bibinfo {author} {\bibfnamefont {K.}~\bibnamefont {Tang}}, \bibinfo {author} {\bibfnamefont {Y.}~\bibnamefont {Zhai}}, \bibinfo {author} {\bibfnamefont {X.}~\bibnamefont {Nie}}, \bibinfo {author} {\bibfnamefont {T.}~\bibnamefont {Xin}}, \bibinfo {author} {\bibfnamefont {J.}~\bibnamefont {Li}},\ and\ \bibinfo {author} {\bibfnamefont {D.}~\bibnamefont {Lu}},\ }\href@noop {} {\bibfield  {journal} {\bibinfo  {journal} {Communications Physics}\ }\textbf {\bibinfo {volume} {7}},\ \bibinfo {pages} {282} (\bibinfo {year} {2024})}\BibitemShut {NoStop}%
\bibitem [{\citenamefont {Liu}\ and\ \citenamefont {Yuan}(2017)}]{LY:17:PRA}%
  \BibitemOpen
  \bibfield  {author} {\bibinfo {author} {\bibfnamefont {J.}~\bibnamefont {Liu}}\ and\ \bibinfo {author} {\bibfnamefont {H.}~\bibnamefont {Yuan}},\ }\href@noop {} {\bibfield  {journal} {\bibinfo  {journal} {Physical Review A}\ }\textbf {\bibinfo {volume} {96}},\ \bibinfo {pages} {012117} (\bibinfo {year} {2017})}\BibitemShut {NoStop}%
\bibitem [{\citenamefont {Pang}\ and\ \citenamefont {Jordan}(2017)}]{PJ:17:Nature}%
  \BibitemOpen
  \bibfield  {author} {\bibinfo {author} {\bibfnamefont {S.}~\bibnamefont {Pang}}\ and\ \bibinfo {author} {\bibfnamefont {A.~N.}\ \bibnamefont {Jordan}},\ }\href@noop {} {\bibfield  {journal} {\bibinfo  {journal} {Nature communications}\ }\textbf {\bibinfo {volume} {8}},\ \bibinfo {pages} {14695} (\bibinfo {year} {2017})}\BibitemShut {NoStop}%
\bibitem [{\citenamefont {Xu}\ \emph {et~al.}(2019)\citenamefont {Xu}, \citenamefont {Li}, \citenamefont {Liu}, \citenamefont {Wang}, \citenamefont {Yuan},\ and\ \citenamefont {Wang}}]{XLL:19:NJPQI}%
  \BibitemOpen
  \bibfield  {author} {\bibinfo {author} {\bibfnamefont {H.}~\bibnamefont {Xu}}, \bibinfo {author} {\bibfnamefont {J.}~\bibnamefont {Li}}, \bibinfo {author} {\bibfnamefont {L.}~\bibnamefont {Liu}}, \bibinfo {author} {\bibfnamefont {Y.}~\bibnamefont {Wang}}, \bibinfo {author} {\bibfnamefont {H.}~\bibnamefont {Yuan}},\ and\ \bibinfo {author} {\bibfnamefont {X.}~\bibnamefont {Wang}},\ }\href@noop {} {\bibfield  {journal} {\bibinfo  {journal} {npj Quantum Information}\ }\textbf {\bibinfo {volume} {5}},\ \bibinfo {pages} {82} (\bibinfo {year} {2019})}\BibitemShut {NoStop}%
\bibitem [{\citenamefont {Yang}\ \emph {et~al.}(2022)\citenamefont {Yang}, \citenamefont {Pang}, \citenamefont {Chen}, \citenamefont {Jordan},\ and\ \citenamefont {Del~Campo}}]{YPC:22:PRL}%
  \BibitemOpen
  \bibfield  {author} {\bibinfo {author} {\bibfnamefont {J.}~\bibnamefont {Yang}}, \bibinfo {author} {\bibfnamefont {S.}~\bibnamefont {Pang}}, \bibinfo {author} {\bibfnamefont {Z.}~\bibnamefont {Chen}}, \bibinfo {author} {\bibfnamefont {A.~N.}\ \bibnamefont {Jordan}},\ and\ \bibinfo {author} {\bibfnamefont {A.}~\bibnamefont {Del~Campo}},\ }\href@noop {} {\bibfield  {journal} {\bibinfo  {journal} {Physical Review Letters}\ }\textbf {\bibinfo {volume} {128}},\ \bibinfo {pages} {160505} (\bibinfo {year} {2022})}\BibitemShut {NoStop}%
\bibitem [{\citenamefont {Zhai}\ \emph {et~al.}(2023)\citenamefont {Zhai}, \citenamefont {Yang}, \citenamefont {Tang}, \citenamefont {Long}, \citenamefont {Nie}, \citenamefont {Xin}, \citenamefont {Lu},\ and\ \citenamefont {Li}}]{ZYKX:23:PRA}%
  \BibitemOpen
  \bibfield  {author} {\bibinfo {author} {\bibfnamefont {Y.}~\bibnamefont {Zhai}}, \bibinfo {author} {\bibfnamefont {X.}~\bibnamefont {Yang}}, \bibinfo {author} {\bibfnamefont {K.}~\bibnamefont {Tang}}, \bibinfo {author} {\bibfnamefont {X.}~\bibnamefont {Long}}, \bibinfo {author} {\bibfnamefont {X.}~\bibnamefont {Nie}}, \bibinfo {author} {\bibfnamefont {T.}~\bibnamefont {Xin}}, \bibinfo {author} {\bibfnamefont {D.}~\bibnamefont {Lu}},\ and\ \bibinfo {author} {\bibfnamefont {J.}~\bibnamefont {Li}},\ }\href@noop {} {\bibfield  {journal} {\bibinfo  {journal} {Physical Review A}\ }\textbf {\bibinfo {volume} {107}},\ \bibinfo {pages} {022602} (\bibinfo {year} {2023})}\BibitemShut {NoStop}%
\bibitem [{\citenamefont {Nichols}\ \emph {et~al.}(2016)\citenamefont {Nichols}, \citenamefont {Bromley}, \citenamefont {Correa},\ and\ \citenamefont {Adesso}}]{NBCA:16:PRA}%
  \BibitemOpen
  \bibfield  {author} {\bibinfo {author} {\bibfnamefont {R.}~\bibnamefont {Nichols}}, \bibinfo {author} {\bibfnamefont {T.~R.}\ \bibnamefont {Bromley}}, \bibinfo {author} {\bibfnamefont {L.~A.}\ \bibnamefont {Correa}},\ and\ \bibinfo {author} {\bibfnamefont {G.}~\bibnamefont {Adesso}},\ }\href@noop {} {\bibfield  {journal} {\bibinfo  {journal} {Phys. Rev. A}\ }\textbf {\bibinfo {volume} {94}},\ \bibinfo {pages} {042101} (\bibinfo {year} {2016})}\BibitemShut {NoStop}%
\bibitem [{\citenamefont {Wang}\ \emph {et~al.}(2018)\citenamefont {Wang}, \citenamefont {Wang}, \citenamefont {Zhan}, \citenamefont {Bian}, \citenamefont {Li}, \citenamefont {Sanders},\ and\ \citenamefont {Xue}}]{WWZB:18:PRA}%
  \BibitemOpen
  \bibfield  {author} {\bibinfo {author} {\bibfnamefont {K.}~\bibnamefont {Wang}}, \bibinfo {author} {\bibfnamefont {X.}~\bibnamefont {Wang}}, \bibinfo {author} {\bibfnamefont {X.}~\bibnamefont {Zhan}}, \bibinfo {author} {\bibfnamefont {Z.}~\bibnamefont {Bian}}, \bibinfo {author} {\bibfnamefont {J.}~\bibnamefont {Li}}, \bibinfo {author} {\bibfnamefont {B.~C.}\ \bibnamefont {Sanders}},\ and\ \bibinfo {author} {\bibfnamefont {P.}~\bibnamefont {Xue}},\ }\href@noop {} {\bibfield  {journal} {\bibinfo  {journal} {Physical Review A}\ }\textbf {\bibinfo {volume} {97}},\ \bibinfo {pages} {042112} (\bibinfo {year} {2018})}\BibitemShut {NoStop}%
\bibitem [{\citenamefont {D\"ur}\ \emph {et~al.}(2014)\citenamefont {D\"ur}, \citenamefont {Skotiniotis}, \citenamefont {Fr\"owis},\ and\ \citenamefont {Kraus}}]{DSF:14:PRL}%
  \BibitemOpen
  \bibfield  {author} {\bibinfo {author} {\bibfnamefont {W.}~\bibnamefont {D\"ur}}, \bibinfo {author} {\bibfnamefont {M.}~\bibnamefont {Skotiniotis}}, \bibinfo {author} {\bibfnamefont {F.}~\bibnamefont {Fr\"owis}},\ and\ \bibinfo {author} {\bibfnamefont {B.}~\bibnamefont {Kraus}},\ }\href@noop {} {\bibfield  {journal} {\bibinfo  {journal} {Phys. Rev. Lett.}\ }\textbf {\bibinfo {volume} {112}},\ \bibinfo {pages} {080801} (\bibinfo {year} {2014})}\BibitemShut {NoStop}%
\bibitem [{\citenamefont {Shettell}\ \emph {et~al.}(2021)\citenamefont {Shettell}, \citenamefont {Munro}, \citenamefont {Markham},\ and\ \citenamefont {Nemoto}}]{SMM:21:NJP}%
  \BibitemOpen
  \bibfield  {author} {\bibinfo {author} {\bibfnamefont {N.}~\bibnamefont {Shettell}}, \bibinfo {author} {\bibfnamefont {W.~J.}\ \bibnamefont {Munro}}, \bibinfo {author} {\bibfnamefont {D.}~\bibnamefont {Markham}},\ and\ \bibinfo {author} {\bibfnamefont {K.}~\bibnamefont {Nemoto}},\ }\href@noop {} {\bibfield  {journal} {\bibinfo  {journal} {New Journal of Physics}\ }\textbf {\bibinfo {volume} {23}},\ \bibinfo {pages} {043038} (\bibinfo {year} {2021})}\BibitemShut {NoStop}%
\bibitem [{\citenamefont {Cai}\ \emph {et~al.}(2023)\citenamefont {Cai}, \citenamefont {Babbush}, \citenamefont {Benjamin}, \citenamefont {Endo}, \citenamefont {Huggins}, \citenamefont {Li}, \citenamefont {McClean},\ and\ \citenamefont {O’Brien}}]{CBBEH:23:RMP}%
  \BibitemOpen
  \bibfield  {author} {\bibinfo {author} {\bibfnamefont {Z.}~\bibnamefont {Cai}}, \bibinfo {author} {\bibfnamefont {R.}~\bibnamefont {Babbush}}, \bibinfo {author} {\bibfnamefont {S.~C.}\ \bibnamefont {Benjamin}}, \bibinfo {author} {\bibfnamefont {S.}~\bibnamefont {Endo}}, \bibinfo {author} {\bibfnamefont {W.~J.}\ \bibnamefont {Huggins}}, \bibinfo {author} {\bibfnamefont {Y.}~\bibnamefont {Li}}, \bibinfo {author} {\bibfnamefont {J.~R.}\ \bibnamefont {McClean}},\ and\ \bibinfo {author} {\bibfnamefont {T.~E.}\ \bibnamefont {O’Brien}},\ }\href@noop {} {\bibfield  {journal} {\bibinfo  {journal} {Reviews of Modern Physics}\ }\textbf {\bibinfo {volume} {95}},\ \bibinfo {pages} {045005} (\bibinfo {year} {2023})}\BibitemShut {NoStop}%
\bibitem [{\citenamefont {Demkowicz-Dobrza\ifmmode~\acute{n}\else \'{n}\fi{}ski}\ \emph {et~al.}(2017)\citenamefont {Demkowicz-Dobrza\ifmmode~\acute{n}\else \'{n}\fi{}ski}, \citenamefont {Czajkowski},\ and\ \citenamefont {Sekatski}}]{DCS:17:PRX}%
  \BibitemOpen
  \bibfield  {author} {\bibinfo {author} {\bibfnamefont {R.}~\bibnamefont {Demkowicz-Dobrza\ifmmode~\acute{n}\else \'{n}\fi{}ski}}, \bibinfo {author} {\bibfnamefont {J.}~\bibnamefont {Czajkowski}},\ and\ \bibinfo {author} {\bibfnamefont {P.}~\bibnamefont {Sekatski}},\ }\href@noop {} {\bibfield  {journal} {\bibinfo  {journal} {Phys. Rev. X}\ }\textbf {\bibinfo {volume} {7}},\ \bibinfo {pages} {041009} (\bibinfo {year} {2017})}\BibitemShut {NoStop}%
\bibitem [{\citenamefont {Huang}\ \emph {et~al.}(2016)\citenamefont {Huang}, \citenamefont {Macchiavello},\ and\ \citenamefont {Maccone}}]{HMM:16:PRA}%
  \BibitemOpen
  \bibfield  {author} {\bibinfo {author} {\bibfnamefont {Z.}~\bibnamefont {Huang}}, \bibinfo {author} {\bibfnamefont {C.}~\bibnamefont {Macchiavello}},\ and\ \bibinfo {author} {\bibfnamefont {L.}~\bibnamefont {Maccone}},\ }\href@noop {} {\bibfield  {journal} {\bibinfo  {journal} {Physical Review A}\ }\textbf {\bibinfo {volume} {94}},\ \bibinfo {pages} {012101} (\bibinfo {year} {2016})}\BibitemShut {NoStop}%
\bibitem [{\citenamefont {Smirne}\ \emph {et~al.}(2016)\citenamefont {Smirne}, \citenamefont {Kołodyński}, \citenamefont {Huelga},\ and\ \citenamefont {Demkowicz-Dobrzański}}]{SKHD:16:PRL}%
  \BibitemOpen
  \bibfield  {author} {\bibinfo {author} {\bibfnamefont {A.}~\bibnamefont {Smirne}}, \bibinfo {author} {\bibfnamefont {J.}~\bibnamefont {Kołodyński}}, \bibinfo {author} {\bibfnamefont {S.~F.}\ \bibnamefont {Huelga}},\ and\ \bibinfo {author} {\bibfnamefont {R.}~\bibnamefont {Demkowicz-Dobrzański}},\ }\href@noop {} {\bibfield  {journal} {\bibinfo  {journal} {Phys. Rev. Lett.}\ }\textbf {\bibinfo {volume} {116}},\ \bibinfo {pages} {120801} (\bibinfo {year} {2016})}\BibitemShut {NoStop}%
\bibitem [{\citenamefont {Haase}\ \emph {et~al.}(2018)\citenamefont {Haase}, \citenamefont {Smirne}, \citenamefont {Kołodyński}, \citenamefont {Demkowicz-Dobrzański},\ and\ \citenamefont {Huelga}}]{HSK:18:NJP}%
  \BibitemOpen
  \bibfield  {author} {\bibinfo {author} {\bibfnamefont {J.~F.}\ \bibnamefont {Haase}}, \bibinfo {author} {\bibfnamefont {A.}~\bibnamefont {Smirne}}, \bibinfo {author} {\bibfnamefont {J.}~\bibnamefont {Kołodyński}}, \bibinfo {author} {\bibfnamefont {R.}~\bibnamefont {Demkowicz-Dobrzański}},\ and\ \bibinfo {author} {\bibfnamefont {S.~F.}\ \bibnamefont {Huelga}},\ }\href@noop {} {\bibfield  {journal} {\bibinfo  {journal} {New Journal of Physics}\ }\textbf {\bibinfo {volume} {20}},\ \bibinfo {pages} {053009} (\bibinfo {year} {2018})}\BibitemShut {NoStop}%
\bibitem [{\citenamefont {Merkli}(2022)}]{MM:22:Quantum}%
  \BibitemOpen
  \bibfield  {author} {\bibinfo {author} {\bibfnamefont {M.}~\bibnamefont {Merkli}},\ }\href@noop {} {\bibfield  {journal} {\bibinfo  {journal} {{Quantum}}\ }\textbf {\bibinfo {volume} {6}},\ \bibinfo {pages} {616} (\bibinfo {year} {2022})}\BibitemShut {NoStop}%
\bibitem [{\citenamefont {Khaneja}\ \emph {et~al.}(2005)\citenamefont {Khaneja}, \citenamefont {Reiss}, \citenamefont {Kehlet}, \citenamefont {Schulte-Herbrüggen},\ and\ \citenamefont {Glaser}}]{KRKH:05:JMR}%
  \BibitemOpen
  \bibfield  {author} {\bibinfo {author} {\bibfnamefont {N.}~\bibnamefont {Khaneja}}, \bibinfo {author} {\bibfnamefont {T.}~\bibnamefont {Reiss}}, \bibinfo {author} {\bibfnamefont {C.}~\bibnamefont {Kehlet}}, \bibinfo {author} {\bibfnamefont {T.}~\bibnamefont {Schulte-Herbrüggen}},\ and\ \bibinfo {author} {\bibfnamefont {S.~J.}\ \bibnamefont {Glaser}},\ }\href@noop {} {\bibfield  {journal} {\bibinfo  {journal} {Journal of Magnetic Resonance}\ }\textbf {\bibinfo {volume} {172}},\ \bibinfo {pages} {296} (\bibinfo {year} {2005})}\BibitemShut {NoStop}%
\bibitem [{\citenamefont {Zhang}\ \emph {et~al.}(2022)\citenamefont {Zhang}, \citenamefont {Yu}, \citenamefont {Yuan}, \citenamefont {Wang}, \citenamefont {Demkowicz-Dobrza\ifmmode~\acute{n}\else \'{n}\fi{}ski},\ and\ \citenamefont {Liu}}]{ZYYW:22:PRR}%
  \BibitemOpen
  \bibfield  {author} {\bibinfo {author} {\bibfnamefont {M.}~\bibnamefont {Zhang}}, \bibinfo {author} {\bibfnamefont {H.-M.}\ \bibnamefont {Yu}}, \bibinfo {author} {\bibfnamefont {H.}~\bibnamefont {Yuan}}, \bibinfo {author} {\bibfnamefont {X.}~\bibnamefont {Wang}}, \bibinfo {author} {\bibfnamefont {R.}~\bibnamefont {Demkowicz-Dobrza\ifmmode~\acute{n}\else \'{n}\fi{}ski}},\ and\ \bibinfo {author} {\bibfnamefont {J.}~\bibnamefont {Liu}},\ }\href@noop {} {\bibfield  {journal} {\bibinfo  {journal} {Phys. Rev. Res.}\ }\textbf {\bibinfo {volume} {4}},\ \bibinfo {pages} {043057} (\bibinfo {year} {2022})}\BibitemShut {NoStop}%
\bibitem [{\citenamefont {Szilva}\ \emph {et~al.}(2023)\citenamefont {Szilva}, \citenamefont {Kvashnin}, \citenamefont {Stepanov}, \citenamefont {Nordstr{\"o}m}, \citenamefont {Eriksson}, \citenamefont {Lichtenstein},\ and\ \citenamefont {Katsnelson}}]{SKS:23:RMP}%
  \BibitemOpen
  \bibfield  {author} {\bibinfo {author} {\bibfnamefont {A.}~\bibnamefont {Szilva}}, \bibinfo {author} {\bibfnamefont {Y.}~\bibnamefont {Kvashnin}}, \bibinfo {author} {\bibfnamefont {E.~A.}\ \bibnamefont {Stepanov}}, \bibinfo {author} {\bibfnamefont {L.}~\bibnamefont {Nordstr{\"o}m}}, \bibinfo {author} {\bibfnamefont {O.}~\bibnamefont {Eriksson}}, \bibinfo {author} {\bibfnamefont {A.~I.}\ \bibnamefont {Lichtenstein}},\ and\ \bibinfo {author} {\bibfnamefont {M.~I.}\ \bibnamefont {Katsnelson}},\ }\href@noop {} {\bibfield  {journal} {\bibinfo  {journal} {Reviews of Modern Physics}\ }\textbf {\bibinfo {volume} {95}},\ \bibinfo {pages} {035004} (\bibinfo {year} {2023})}\BibitemShut {NoStop}%
\bibitem [{\citenamefont {Ali}\ \emph {et~al.}(2024{\natexlab{a}})\citenamefont {Ali}, \citenamefont {Al-Kuwari}, \citenamefont {Rahim}, \citenamefont {Ghominejad}, \citenamefont {Ali},\ and\ \citenamefont {Haddadi}}]{AAR:APB:24}%
  \BibitemOpen
  \bibfield  {author} {\bibinfo {author} {\bibfnamefont {A.}~\bibnamefont {Ali}}, \bibinfo {author} {\bibfnamefont {S.}~\bibnamefont {Al-Kuwari}}, \bibinfo {author} {\bibfnamefont {M.}~\bibnamefont {Rahim}}, \bibinfo {author} {\bibfnamefont {M.}~\bibnamefont {Ghominejad}}, \bibinfo {author} {\bibfnamefont {H.}~\bibnamefont {Ali}},\ and\ \bibinfo {author} {\bibfnamefont {S.}~\bibnamefont {Haddadi}},\ }\href@noop {} {\bibfield  {journal} {\bibinfo  {journal} {Applied Physics B}\ }\textbf {\bibinfo {volume} {130}},\ \bibinfo {pages} {177} (\bibinfo {year} {2024}{\natexlab{a}})}\BibitemShut {NoStop}%
\bibitem [{\citenamefont {Ali}\ \emph {et~al.}(2024{\natexlab{b}})\citenamefont {Ali}, \citenamefont {Al-Kuwari}, \citenamefont {Hussain}, \citenamefont {Byrnes}, \citenamefont {Rahim}, \citenamefont {Quach}, \citenamefont {Ghominejad},\ and\ \citenamefont {Haddadi}}]{AAH:24:PRA}%
  \BibitemOpen
  \bibfield  {author} {\bibinfo {author} {\bibfnamefont {A.}~\bibnamefont {Ali}}, \bibinfo {author} {\bibfnamefont {S.}~\bibnamefont {Al-Kuwari}}, \bibinfo {author} {\bibfnamefont {M.}~\bibnamefont {Hussain}}, \bibinfo {author} {\bibfnamefont {T.}~\bibnamefont {Byrnes}}, \bibinfo {author} {\bibfnamefont {M.}~\bibnamefont {Rahim}}, \bibinfo {author} {\bibfnamefont {J.~Q.}\ \bibnamefont {Quach}}, \bibinfo {author} {\bibfnamefont {M.}~\bibnamefont {Ghominejad}},\ and\ \bibinfo {author} {\bibfnamefont {S.}~\bibnamefont {Haddadi}},\ }\href@noop {} {\bibfield  {journal} {\bibinfo  {journal} {Physical Review A}\ }\textbf {\bibinfo {volume} {110}},\ \bibinfo {pages} {052404} (\bibinfo {year} {2024}{\natexlab{b}})}\BibitemShut {NoStop}%
\bibitem [{\citenamefont {Brask}\ \emph {et~al.}(2015)\citenamefont {Brask}, \citenamefont {Chaves},\ and\ \citenamefont {Kołodyński}}]{BCK:15:PRX}%
  \BibitemOpen
  \bibfield  {author} {\bibinfo {author} {\bibfnamefont {J.~B.}\ \bibnamefont {Brask}}, \bibinfo {author} {\bibfnamefont {R.}~\bibnamefont {Chaves}},\ and\ \bibinfo {author} {\bibfnamefont {J.}~\bibnamefont {Kołodyński}},\ }\href@noop {} {\bibfield  {journal} {\bibinfo  {journal} {Phys. Rev. X}\ }\textbf {\bibinfo {volume} {5}},\ \bibinfo {pages} {031010} (\bibinfo {year} {2015})}\BibitemShut {NoStop}%
\bibitem [{\citenamefont {Koch}\ \emph {et~al.}(2022)\citenamefont {Koch}, \citenamefont {Boscain}, \citenamefont {Calarco}, \citenamefont {Dirr}, \citenamefont {Filipp}, \citenamefont {Glaser}, \citenamefont {Kosloff}, \citenamefont {Montangero}, \citenamefont {Schulte-Herbr{\"u}ggen}, \citenamefont {Sugny} \emph {et~al.}}]{KBCD:22:EPJ}%
  \BibitemOpen
  \bibfield  {author} {\bibinfo {author} {\bibfnamefont {C.~P.}\ \bibnamefont {Koch}}, \bibinfo {author} {\bibfnamefont {U.}~\bibnamefont {Boscain}}, \bibinfo {author} {\bibfnamefont {T.}~\bibnamefont {Calarco}}, \bibinfo {author} {\bibfnamefont {G.}~\bibnamefont {Dirr}}, \bibinfo {author} {\bibfnamefont {S.}~\bibnamefont {Filipp}}, \bibinfo {author} {\bibfnamefont {S.~J.}\ \bibnamefont {Glaser}}, \bibinfo {author} {\bibfnamefont {R.}~\bibnamefont {Kosloff}}, \bibinfo {author} {\bibfnamefont {S.}~\bibnamefont {Montangero}}, \bibinfo {author} {\bibfnamefont {T.}~\bibnamefont {Schulte-Herbr{\"u}ggen}}, \bibinfo {author} {\bibfnamefont {D.}~\bibnamefont {Sugny}}, \emph {et~al.},\ }\href@noop {} {\bibfield  {journal} {\bibinfo  {journal} {EPJ Quantum Technology}\ }\textbf {\bibinfo {volume} {9}},\ \bibinfo {pages} {19} (\bibinfo {year} {2022})}\BibitemShut {NoStop}%
\bibitem [{\citenamefont {Lu}\ \emph {et~al.}(2024)\citenamefont {Lu}, \citenamefont {Joshi}, \citenamefont {San~Dinh},\ and\ \citenamefont {Koch}}]{LJSK:24:JPC}%
  \BibitemOpen
  \bibfield  {author} {\bibinfo {author} {\bibfnamefont {Y.}~\bibnamefont {Lu}}, \bibinfo {author} {\bibfnamefont {S.}~\bibnamefont {Joshi}}, \bibinfo {author} {\bibfnamefont {V.}~\bibnamefont {San~Dinh}},\ and\ \bibinfo {author} {\bibfnamefont {J.}~\bibnamefont {Koch}},\ }\href@noop {} {\bibfield  {journal} {\bibinfo  {journal} {Journal of Physics Communications}\ }\textbf {\bibinfo {volume} {8}},\ \bibinfo {pages} {025002} (\bibinfo {year} {2024})}\BibitemShut {NoStop}%
\bibitem [{\citenamefont {Wang}\ \emph {et~al.}(2023)\citenamefont {Wang}, \citenamefont {Okyay}, \citenamefont {Kumar},\ and\ \citenamefont {Wong}}]{WOKW:23:QS}%
  \BibitemOpen
  \bibfield  {author} {\bibinfo {author} {\bibfnamefont {X.}~\bibnamefont {Wang}}, \bibinfo {author} {\bibfnamefont {M.~S.}\ \bibnamefont {Okyay}}, \bibinfo {author} {\bibfnamefont {A.}~\bibnamefont {Kumar}},\ and\ \bibinfo {author} {\bibfnamefont {B.~M.}\ \bibnamefont {Wong}},\ }\href@noop {} {\bibfield  {journal} {\bibinfo  {journal} {AVS Quantum Science}\ }\textbf {\bibinfo {volume} {5}} (\bibinfo {year} {2023})}\BibitemShut {NoStop}%
\bibitem [{\citenamefont {Rodr{\'\i}guez-Borb{\'o}n}\ \emph {et~al.}(2024)\citenamefont {Rodr{\'\i}guez-Borb{\'o}n}, \citenamefont {Wang}, \citenamefont {Di{\'e}guez}, \citenamefont {Ibrahim},\ and\ \citenamefont {Wong}}]{RWD:24:CPC}%
  \BibitemOpen
  \bibfield  {author} {\bibinfo {author} {\bibfnamefont {J.~M.}\ \bibnamefont {Rodr{\'\i}guez-Borb{\'o}n}}, \bibinfo {author} {\bibfnamefont {X.}~\bibnamefont {Wang}}, \bibinfo {author} {\bibfnamefont {A.~P.}\ \bibnamefont {Di{\'e}guez}}, \bibinfo {author} {\bibfnamefont {K.~Z.}\ \bibnamefont {Ibrahim}},\ and\ \bibinfo {author} {\bibfnamefont {B.~M.}\ \bibnamefont {Wong}},\ }\href@noop {} {\bibfield  {journal} {\bibinfo  {journal} {Computer Physics Communications}\ ,\ \bibinfo {pages} {109403}} (\bibinfo {year} {2024})}\BibitemShut {NoStop}%
\end{thebibliography}%

\end{document}